\documentclass[lettersize,onecolumn]{IEEEtran}

\usepackage{amsmath,amsfonts}
\usepackage{algorithmic}
\usepackage{algorithm}
\usepackage{array}
\usepackage[caption=false,font=normalsize,labelfont=sf,textfont=sf]{subfig}
\usepackage{textcomp}
\usepackage{stfloats}
\usepackage{url}
\usepackage{verbatim}
\usepackage{graphicx}
\usepackage{cite}
\usepackage{xfrac}
\usepackage{xcolor}
\usepackage{dsfont}
\usepackage{makecell}
\usepackage{multirow}
\usepackage{hyperref}
\usepackage{amsthm}
\usepackage{multicol}

\usepackage{balance}

\theoremstyle{plain}
\newtheorem{theorem}{Theorem}
\newtheorem{lemma}[theorem]{Lemma}
\newtheorem{corollary}[theorem]{Corollary}
\newtheorem{proposition}{Proposition}
\newtheorem{claim}{Claim}
\theoremstyle{definition}
\newtheorem{defn}{Definition}
\newtheorem{rem}[theorem]{Remark}

\newcommand{\openone}{\mathds{1}}

\newcommand{\Vc}{\mathcal{V}}

\newcommand{\Oc}{\mathcal{O}}

\newcommand{\var}{\text{var}}
\newcommand{\cov}{\text{cov}}

\newcommand{\Xc}{\mathcal{X}}

\newcommand{\Pc}{\mathcal{P}}

\newcommand{\phat}{\hat{p}}

\newcommand\numberthis{\addtocounter{equation}{1}\tag{\theequation}}
\newcommand{\Pbar}{\bar{\mathbf{P}}}
\newcommand{\Phat}{\hat{P}}
\newcommand{\sigmab}{\bar{\sigma}}
\newcommand{\Qbar}{\bar{\mathbf{Q}}}

\begin{document}

\title{Model non-collapse: Minimax bounds for recursive discrete distribution estimation
	\thanks{The work in this paper was supported in part by the Swiss National Science Foundation under Grant 200364.}%
	}

\author{Millen Kanabar %
	and Michael Gastpar\\ %
	School of Computer and Communication Sciences, EPFL\thanks{Email: \href{mailto:millen.kanabar@epfl.ch}{millen.kanabar@epfl.ch} and \href{mailto:michael.gastpar@epfl.ch}{michael.gastpar@epfl.ch}}}%

\maketitle

\begin{abstract}
Learning discrete distributions from i.i.d. samples is a well-understood problem. However, advances in generative machine learning prompt an interesting new, non-i.i.d. setting: after receiving a certain number of samples, an estimated distribution is fixed, and samples from this estimate are drawn and introduced into the sample corpus, undifferentiated from real samples. Subsequent generations of estimators now face contaminated environments, a scenario referred to in the machine learning literature as self-consumption. 

Empirically, it has been observed that models in fully synthetic self-consuming loops collapse---their performance deteriorates with each batch of training---but accumulating data has been shown to prevent complete degeneration. This, in turn, begs the question: What happens when fresh real samples \textit{are} added at every stage? 
In this paper, we study the minimax loss of self-consuming discrete distribution estimation in such loops. We show that even when model collapse is consciously averted, the ratios between the minimax losses with and without source information can grow unbounded as the batch size increases.

In the data accumulation setting, where all batches of samples are available for estimation, we provide minimax lower bounds and upper bounds that are order-optimal under mild conditions for the expected $\ell_2^2$ and $\ell_1$ losses at every stage. We provide conditions for regimes where there is a strict gap in the convergence rates compared to the corresponding oracle-assisted minimax loss where real and synthetic samples are differentiated, and provide examples where this gap is easily observed. We also provide a lower bound on the minimax loss in the data replacement setting, where only the latest batch of samples is available, and use it to find a lower bound for the worst-case loss for bounded estimate trajectories.
\end{abstract}

\section{Introduction}
\IEEEPARstart{T}{he} problem of estimating distributions from their samples arises naturally in various contexts, and has been studied in depth in the literature. The exact minimax loss of discrete distribution estimation under the $\ell_2^2$ metric, among others, was found in \cite{kamath_learning_2015}, and was studied under $\ell_1$ metric e.g. in \cite{chan_learning_2013, han_minimax_2015, kamath_learning_2015}. Evidently, the results of these estimations are often themselves valid distributions.
These estimates can then themselves be used to generate new samples in the same alphabet. 

In cases where distribution estimation is performed in batches, samples generated from estimates at some stage can be reintroduced into subsequent batches of `real' samples, thereby affecting future performance. This phenomenon has been widely observed, for example, in the context of generative machine learning, where distributions are estimated primarily to produce new samples, in part because it is significantly cheaper to generate synthetic samples. This has also led to fears that synthetic samples might significantly contaminate real databases such as the Internet \cite{mahadevan_this_2023}. It has been observed empirically e.g. in \cite{shumailov_ai_2024, alemohammad2024selfconsuming} that this process, without external mitigation, leads to `model collapse': worsening performance with every iteration where samples from previous stages are introduced into the training pool. 

The self-referential nature of the sampling process causes the distributions of later samples to depend non-trivially on the previous batches, and thus prevents the direct use of already existing results from the distribution estimation literature. In this paper, we study the properties of such sequences of estimates under a minimax lens. 

\begin{figure}
	\begin{center}
		\includegraphics[width=0.6\linewidth]{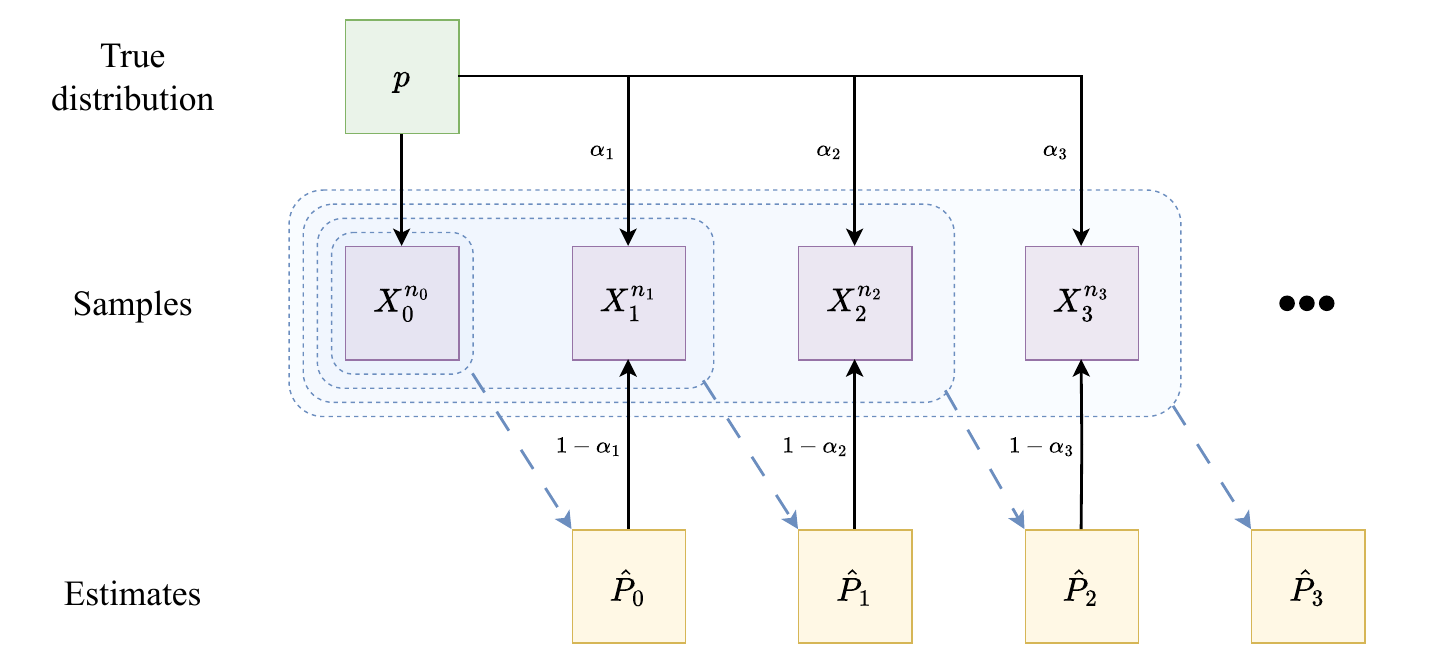}
		\caption{The estimation workflow in the data accumulation setting; solid arrows indicate sampling (with the corresponding sampling probability indicated next to them), whereas dotted arrows indicate estimation. Under the data replacement setting, the corpus used for estimation only contains the latest batch of samples. }
		\label{fig:self-consumption}
	\end{center}
\end{figure}

A sketch of the data production workflow considered in our paper is shown in Figure \ref{fig:self-consumption}. A crucial aspect of the model is that the corpus does not include the identities of the source of the samples, i.e.\ whether they are real or synthetic. In the data accumulation setting, new samples are continually added to the corpus at every stage, whereas it consists of only fresh samples in the data replacement setting. This mirrors quite a few variants considered in the generative machine learning literature. In this paper, we assume that each sample is drawn from either the true underlying distribution or the previous estimate based on the outcome of a coin flip, independent of every other sample, with stage-dependent bias $\alpha_t$ which is assumed to be known to the estimator.

\subsection{Contributions}
In this work, we pose the multi-stage distribution estimation problem with self-consumption of samples, which has not been studied in the minimax setting before in the literature. We analyze the minimax loss in terms of the sequences $n_t$, denoting the total number of samples in batch $t$, and $\alpha_t$, the probability with which each sample in batch $t$ is drawn from the real distribution.
We consider the oracle-assisted minimax loss---where the source distribution of each sample is provided to the estimator by an oracle---as a baseline. It is perhaps unsurprising that the oracle-assisted minimax loss at stage $t$ is $\Theta(\sfrac{1}{\sum_{i=0}^t n_i\alpha_i})$, since the expected number of real samples seen after $t$ batches is precisely $\sum_{i=0}^t n_i\alpha_i$.

We provide the first (to the best of our knowledge) information-theoretic lower bounds on the $\ell_2^2$ and $\ell_1$ losses at each stage in terms of properties of the estimates at previous stages in Theorem~\ref{thm:lower-bound} (for the data accumulation workflow) and Corollary \ref{cor:lower-bound-replacement} (for the data replacement workflow).
We also propose a sequence of estimators that is order-optimal under certain regimes with stage-wise performance guarantees in Theorem~\ref{thm:upper-bound}. We show conditions where there are gaps between the upper bound and lower bounds on the minimax loss and the oracle-assisted minimax loss the data accumulation workflow (Propositions \ref{prop:matching} and \ref{prop:non-matching}) and provide examples where these conditions are satisfied. We briefly discuss the performance of estimators that do not account for the introduction of synthetic samples and show that results shown in the Model Collapse literature, where collapse is avoided by simply following the accumulate workflow still hold in our setting (Corollary \ref{cor:upper-bound-unprocessed}).
Finally, we use our lower bound for the data replacement workflow to find a time-independent lower bound on the worst-case (in ground truth distribution) loss given a time-independent upper bound.

We find that the minimax loss stays within a constant factor of the oracle-assisted one when $\alpha_t$ is lower bounded by a constant. When $\alpha_t$ decays to $0$, i.e. when synthetic sampling outpaces real sampling, the effective sample size of batch $t$ is proportional to $n_t\alpha_t^2$ (as opposed to simply $n_t\alpha_t$ for the oracle-assisted setting) in most practically conceivable conditions. As a consequence, the ratio of the minimax loss to the oracle-assisted minimax loss can grow unbounded with $t$ (e.g. Claim \ref{ex:matching}). We specify the conditions for this regime in Proposition~\ref{prop:matching}; the basic requirement is that $\alpha_t$ does not decay too quickly. 

When $\alpha_t$ does decay quickly, an error term starts dominating the lower bound, and there is a gap between the upper and lower bounds. However, under mild conditions, there is also a gap between the lower bound and oracle-assisted loss (Claim \ref{ex:non-matching}): the effective sample size for the lower bound becomes proportional to $\sqrt{\sum_{i=0}^t n_i\alpha_i}$; it remains $\sum_{i=0}^t n_i\alpha_i^2$ for the upper bound. This regime is analyzed in detail in Proposition~\ref{prop:non-matching}. 

The various regimes are summarized in Table \ref{tab:regimes}.

\begin{table*}[h]
	\centering
	\caption{The various regimes of non-collapsing discrete distribution estimation under $\ell_2^2$ loss in the accumulation workflow}
	\def\arraystretch{1.6}
	\begin{tabular}{|c|c|c|c|c|}
		\hline
		\multicolumn{2}{|c|}{\textbf{Regime}} & \multirow{2}{*}{\makecell{\textbf{Upper Bound}\\ \textit{(Theorem~\ref{thm:upper-bound})}}} & \multirow{2}{*}{\makecell{\textbf{Lower Bound}\\ \textit{(Theorem~\ref{thm:lower-bound})}}} & \multirow{2}{*}{\makecell{\textbf{Oracle-assisted loss}\\\textit{(Lemma~\ref{lem:oracle-minimax})}}} \\
		\cline{1-2}
		Conditions on $\alpha_t$ & Other conditions& & &  \\
		\hline 
		$\alpha_i = \Omega(1)$& - & \multicolumn{3}{c|}{\makecell{\textit{(Proposition~\ref{prop:basic})}\\Both base and oracle-assisted minimax losses are $\Theta\left(\frac{1}{\sum_{i=0}^tn_i\alpha_i}\right)$}} \\
		\hline
		\multirow{2}{*}{\makecell{$\alpha_t\downarrow0$, but\\  $\alpha_t = \Omega\left(\frac{1}{\sum_{i=0}^{t-1}n_i\alpha_i^2}\right)$}} & -&  \multicolumn{2}{c|}{\makecell{\textit{(Proposition~\ref{prop:matching})} \\ Bounds match, minimax loss is $\Theta\left(\frac{1}{\sum_{i=0}^{t}n_i\alpha_i^2}\right)$}} & \multirow[c]{5}{*}{$\Theta\left(\frac{1}{\sum_{i=0}^t n_i\alpha_i}\right)$} \\
		\cline{2-4}
		& $\frac{1}{\sum_{i=0}^{t}n_i\alpha_i^2} = o\left(\alpha_t\right)$& \multicolumn{2}{c|}{\makecell{Additionally, the ratio of the upper and lower bounds \\ is bounded by a constant independent of $k$ as $t\uparrow\infty$}} & \\
		\cline{1-4}
		\multirow[c]{3}{*}{\makecell{$\alpha_t\downarrow0$ and \\ $\alpha_t = o\left(\frac{1}{\sum_{i=0}^{t-1}n_i\alpha_i^2}\right)$}} & \makecell{$n_t\alpha_t \leq \sum_{i=0}^tn_i\alpha_i^2$, \\$n_t\alpha_t = \Oc\left( \sum_{i=0}^{t-1} n_i\alpha_i\right)$  \\and $\sum_{i=0}^t n_i\alpha_i\uparrow\infty$} & $\Oc\left(\frac{1}{\sum_{i=0}^{t}n_i\alpha_i^2}\right)$ & \multirow{2}{*}{\makecell{\textit{(Proposition~\ref{prop:non-matching})}\\$\Omega\left(\frac{1}{\sqrt{\sum_{i=0}^tn_i\alpha_i}}\right)$}} &  \\
		\cline{2-3}
		& \makecell{$n_t\alpha_t = \Oc\left({\sum_{i=0}^{t-1} n_i\alpha_i}\right)$  \\and $\sum_{i=0}^t n_i\alpha_i\uparrow\infty$} & \multirow{2}{*}{\makecell{\textit{(Corollary \ref{cor:upper-bound-unprocessed})}\\$\Oc\left(1\right)$}} & & \\
		\cline{2-2}\cline{4-5}
		& ${\sum_{i=0}^{t-1} n_i\alpha_i} = o(n_t\alpha_t)$&  & $\Omega\left(\frac{1}{n_t\alpha_t}\right)$ & $\Theta\left(\frac{1}{n_t\alpha_t}\right)$ \\
		\hline
		\multicolumn{2}{|c|}{\makecell{ \textit{(Special case)} \\ $\sum_{i=0}^t n_i\alpha_i = \Oc(1)$}}& \multicolumn{3}{c|}{All minimax losses are $\Theta\left(1\right)$}   \\
		\hline
	\end{tabular}
	\label{tab:regimes}
\end{table*}

A few additional technical assumptions are required for the results shown in the table: these are discussed in the corresponding statements.

\subsection{Related work}

Model collapse in settings containing feedback loops with different mechanisms of mixing synthetic data with old and new real samples was first studied empirically in \cite{shumailov_ai_2024} and \cite{alemohammad2024selfconsuming}. Specifically, the case for discrete distribution estimation in a fully synthetic loop, wherein, at every stage, the estimator only has access to samples generated from the distribution estimated at the previous stage was considered as a motivating example in the analysis of \cite{shumailov_ai_2024}. Further progress centered around collapse in fully synthetic loops in discrete distribution estimation was made in \cite{seddik2024how, suresh2025rate}.

There has been a significant amount of recent work on model collapse in synthetic and mixed data loops. An upper bound on the proportion of synthetic data for in-loop stability of parametric generative models trained on freshly generated (mixed) samples at each stage was provided in \cite{bertrand2024on}. The evolution of parametric models and nonparametric kernel density estimators trained on fully synthetic and mixed data loops was described in \cite{fu_towards_2024}. Model collapse in regression models and model collapse mitigation strategies are discussed in \cite{dohmatob_model_2024, gerstgrasser_is_2024, feng_beyond_2024}; a lower bound on the expected error for regression models that do not account for self-consumption was presented in \cite{dohmatob2025strong}. A stochastic-approximation-style treatment showing eventual degeneration for accumulating loops with no fresh real data is provided in \cite{10886816}. Continuing progress on averting model collapse in data accumulation-style workflows, upper bounds on the estimation error in self-consuming loops where real data remains in the training set are studied in \cite{dey_universality_2024, amin2025escaping, fu2025a, kazdan2025collapse, he_golden_2025}.

A central thread in these works is that the distribution of the data used in later generations of estimates (or training later generations of models) is sampled from functions of the empirical distributions of data seen by earlier generations. This is explicit in the works studying generative machine learning settings, and implicit in works studying model collapse in settings like, for example, regression. While our focus in this work is on the evolution of explicit estimates of the true distribution itself, we believe the results hold relevance for most settings considered in the literature.

We note, however, that our setting differs from these works in our assumption that the estimator can account for the presence of unlabeled synthetic data in the corpus of samples. In contrast, the above works study estimators that act on (treated or untreated) mixed datasets no differently than they would on real datasets. Since our setting also allows for such estimators, lower bounds presented here also function as lower bounds for settings considered in previous model collapse literature. 

Our setting also differs from the various settings in the literature where some of the synthetic samples are drawn from estimates, perhaps of a different distribution than the true distribution, based on independent, unseen data \footnote{e.g. in \cite{jain_scaling_2024}, where a careful weighting of these `surrogate' synthetic samples proves beneficial}. In our setting, all information about the true distribution available to the estimators is contained in the samples that they observe---minimax rates using \textit{labeled} synthetic samples from \mbox{(semi-)independent} estimates remain open.

\subsection{When does a model collapse?}

There is no current consensus in on what \textit{precisely} constitutes model collapse\footnote{A list of definitions of model collapse and further discussion can be found in \cite{schaeffer_position_2025}.}. However, based on the above results, we make the following observations vis-\`a-vis the various definitions in literature:
\subsubsection{`Non-collapse'} This work mainly focuses on cases where enough real samples are available such that the estimates eventually converge to the true distribution, or remain within a small radius around it. This avoids, by assumption, the conditions for model collapse (collapsing variance, unbounded growth of the expected loss etc.) discussed in prior work. The primary difference is perhaps a focus in the early literature on the replace workflow, where sequences of estimates behave essentially like unreinforced random walks. Even so, our analysis of the replace workflow provides an impossibility result for the expected loss, showing that even the best-case loss cannot be too far from the average-case loss in the oracle-annotated setting.
\subsubsection{Change in scaling laws with (eventually) ubiquitous synthetic data} When the mixing factor $\alpha_t$, i.e.\ the probability of sampling the true distribution decreases to $0$, no matter how slowly, we can show gaps between the minimax loss with and without the presence of synthetic samples. This points to a change in scaling laws, and hence \textit{does} constitute model collapse according to the criteria in \cite{dohmatob2025strong}. This is a result of the sampling process itself, and cannot be mitigated without further knowledge of the true distribution, e.g. thorough a (possibly noisy) oracle or via human feedback. In this sense, the introduction of synthetic samples changes the \textit{fundamental hardness} of the estimation problems regardless of the estimation procedures used.
\subsubsection{Unchanged scaling in the presence of persistent fractions of real samples} We also find that such gaps do not exist when $\alpha_t$ is is bounded below by a positive constant, i.e.\ when there is a non-vanishing probability of drawing a real sample in each batch. In such cases, model collapse in the discrete distribution estimation setting \textit{can be avoided entirely} in all accepted senses of the term.
\subsection{Notation}
Deterministic quantities and functions are represented by lowercase Greek and Roman alphabets; random variables are represented by uppercase Roman alphabets. The probability simplex corresponding to the alphabet of size $k$ is denoted as $\Delta_k$. We use $[a:b]$ to denote $\{x\in \mathbb{N}: a\leq x\leq b\}$. The unit vector in with component $j$ equal to $1$ is denoted as $e_j$.

The $j^{\mathrm{th}}$ component of a discrete distribution $P$ or vector $p$ is indexed by square brackets as $P[j]$ and $p[j]$ respectively, and the probability of an event $A$ under distribution $Q$ is denoted as $Q\{A\}$. The first subscript of quantities in sequences denotes the index in the sequence, subsequent ones are used to denote other parameters when needed. The $n$-fold product of a distribution $Q$ is denoted as $Q^n$. 

For $t\geq 0$ and functions $f, g$, we write $g(t) = \Theta(f(t))$ if there exist constants $a_1, a_2 \geq 0$ such that $a_1f(t) \leq g(t) \leq a_2f(t) \ \forall t\geq 0$; $g(t) = \Oc(f(t)) $ if there exists a constant $a_3\geq 0$ such that $g(t) \leq a_3f(t) \ \forall t\geq 0$; $g(t)=\Omega(f(t))$ if there exists a constant $a_4>0$ such that $g(t) \geq a_4 f(t) \forall t\geq 0$; $g(t)=o(f(t))$ if $\lim_{t\uparrow\infty} g(t)/f(t) = 0$. 

\section{Problem Setup}
We consider the problem of recursive discrete distribution estimation over finite alphabets. Let $\Xc:=\{1, 2, \dots, k\}$ be the alphabet of size $k$. A sequence of sample-estimator pairs is defined as follows.
\begin{defn}[Self-consuming estimation]\label{def:self-consuming-seq}
	For a distribution $p\in \Delta_k$, a self-consuming sequence of sample-estimator pairs with sample sizes $(n_t)_{t\geq 0}$ and sampling probabilities $(\alpha_t)_{t\geq 1}$ is a sequence $\left(X_{t}^{n_t}, \phat_{t}(\cdot)\right)_{t\geq 0}$ of samples $X_i^{n_i} \in \Xc^{n_i}$ and estimators $\phat_{i}:\Xc^{n_0}\times\dots\times \Xc^{n_t}\rightarrow \Delta_k$ such that \begin{itemize}
		\item $X_0^{n_0} \sim p^{n_0}$, and
		\item $X_{t+1}^{n_{t+1}} \sim \left(\tilde{P}({p, \Phat_t, \alpha_{t+1}})\right)^{n_{t+1}}$,\\
		where $\tilde{P}({p_1, p_2, \alpha}) := \alpha p_1 + (1-\alpha)p_2$, and $\Phat_t := \phat_t\left(X_0^{n_0}, \dots, X_t^{n_t}\right)$ for every $t\geq 0$.
	\end{itemize} 
	The joint distribution of the random vector $(X_0^{n_0}, \dots, X_t^{n_t})$ is denoted as $\Pbar_{t, ({p, \{n_i, \alpha_i: i\geq 0\}})}$. We drop the sample sizes $n_t$ and mixing probabilities $\alpha_t$ from the subscript when they are obvious from the context. By definition, $\alpha_0 = 1$.
\end{defn}

The definition of this sequence encapsulates the following process: at stage $0$, an estimate $\Phat_0$ is produced using a batch of $n_0$ i.i.d. samples from $p$. At every subsequent stage $t+1$, a batch of $n_{t+1}$ samples is produced, with each sample being distributed as $p$ with probability $\alpha_{t+1}$ and as $\Phat_{t}$ with probability $1-\alpha_{t+1}$. Since this process can be seen as conditional sampling based on the outcome of a biased coin-flip, we refer to the outcome of this flip as the `identity' of the source.

For a sequence of estimators $\phat_t(\cdot)$, we consider the minimax loss under the $\ell_2^2$ and $\ell_1$ losses.
\begin{defn}[Minimax loss]\label{def:minimax-loss}
	For a given loss function $\ell:\Delta_k\times \Delta_k\rightarrow\mathbb{R}_+$, sequences $\{n_i, \alpha_i: i\geq 0\}$ and fixed estimators $\phat_{i}(\cdot), i \in [1:t]$, the minimax loss at stage $t$ is defined as
	\begin{align*}
		r^{\ell}_{t, k} = \inf_{\phat_t}\sup_{p \in \Delta_k} E[\ell(\phat_t(X_0^{n_0}, \dots, X_t^{n_t}), p)]
	\end{align*}
	where the expectation is with respect to the joint distribution $\Pbar_{t, (p, \{n_i, \alpha_i: i\geq 0\})}$. 
\end{defn} 

It is important to note that the minimax loss is defined for the estimator at every stage $t$ while fixing the estimators in previous stages. This follows from the process of designing the estimators: they are selected sequentially, optimized at every stage, without concern for how the samples generated from these estimates might influence future samples and without mechanisms for modifying previously made estimates.

\section{Main Results}
The main contributions of this paper are upper and lower bounds on the minimax loss of the self-consuming sequence at an arbitrary stage $t\geq 0$. Recall that at stage $t$, the probability of drawing a sample from the true distribution $p$ is denoted by $\alpha_t$, the estimated distribution is denoted by $\Phat_t$, and the joint distribution of the samples collected and distributions estimated so far is denoted by $\Pbar_{t, (p)}$.

\subsection{Bounds on the minimax loss}

We are now ready to state the main results of this paper. The lower bound on the minimax loss is given in terms of an error term, defined subsequently.

\begin{theorem}[Minimax lower bound]\label{thm:lower-bound}
	Fix $\phat^{(i)}, i{\in} [0{:} t{-}1]$. 	
	The minimax loss for the self-consuming estimator-sample sequence at stage $t$ under $\ell_2^2$ and $\ell_1$ losses is lower bounded, respectively, as
	\begin{align}
		r^{\ell_2^2}_{t, k} \geq& \frac{\sfrac{1}{512}}{\sum\limits_{i\in [0:t]} n_i\alpha_i^2 \, + h_t({\sfrac{1}{4k}})}\\
		r^{\ell_1}_{t, k} \geq& \sqrt{\frac{\sfrac{k}{4096}}{\sum\limits_{i\in [0:t]} n_i\alpha_i^2 \, + h_t({\sfrac{1}{4k}})}},
	\end{align}whenever $\sum\limits_{i\in [0:t]} n_i\alpha_i^2 \, + h_{t}(\epsilon) \geq k/4$, where $h_t(.)$ is defined as in Definition \ref{def:error-term}. 
\end{theorem}

\begin{defn}[Error term for lower bounds]\label{def:error-term}
	The error term $h_{t}(\epsilon)$ is defined as \begin{align*}
		h_t(\epsilon) := \sum_{i=0}^{t} n_i\alpha_ig_i(\epsilon),
	\end{align*} where \begin{align*}
		g_{i}(\epsilon) &:= \sup_p \max_j \Pbar_{i-1,({p})}\left\{\Phat_{i-1}[j]< p[j]-\epsilon\right\}
	\end{align*}denotes the worst case (component-wise and distribution-wise) probability of an $\epsilon$-error in $\Phat_{i-1}$ under $\Pbar_{i-1,({p})}$ for all $i\neq 0$, and $g_{0}(\epsilon) = 0 \ \forall \epsilon$. 
\end{defn}

\begin{theorem}[Upper bound]\label{thm:upper-bound}
	Let $n_i\alpha_i \leq \sum_{j\in [0:i]} n_j \alpha_j^2$ for every $i \leq t$. Then there exists a sequence of estimators $\phat_{i}, i\geq 0$ for which the worst-case $\ell_2^2$ and $\ell_1$ losses are upper bounded as
	\begin{align}
		\sup_p E[\ell_2^2(p, \phat_t(X_0^{n_0}, \dots, X_t^{n_t}))] \leq& \frac{1-\sfrac{1}{k}}{\sum\limits_{i\in [0:t]} n_i\alpha_i^2} \label{eq:upper-bound-l2-combined}\\
		\sup_p E[\ell_1(p, \phat_t(X_0^{n_0}, \dots, X_t^{n_t}))] \leq& \sqrt{\frac{k-1}{\sum\limits_{i\in [0:t]} n_i\alpha_i^2}}. \label{eq:upper-bound-l1-combined}
	\end{align}
\end{theorem}

We can directly observe that when the error term $h_t(\sfrac{1}{4k})$ is small, the minimax errors are within constant factors of each other; and, as we will see in the baseline, the erasure of the source information leads to a penalty of $\alpha_t$ on the effective sample size of batch $t$ (as compared to when the estimators know which distribution each sample is drawn from). As a corollary of the analysis of Theorem~\ref{thm:upper-bound}, we also describe the worst-case error when the sequence of empirical estimators without further processing is used. 

\begin{corollary}[Unprocessed upper bound]\label{cor:upper-bound-unprocessed}
	For the sequence of estimators $\Phat_{t}^{\mathrm{emp}}:= \phat_{\mathrm{emp}}(X_0^{n_0}, \dots, X_t^{n_t})$, where $\phat_{\mathrm{emp}}(\cdot)$ denotes the empirical estimator of all samples, the worst-case $\ell_2^2$ loss under self-consumption at stage $t$ is bounded above as
	\begin{align}
		\sup_{p \in \Delta_k}E[\ell_2^2(\Phat_t^{\mathrm{emp}}, p)] \leq \left(1-\frac{1}{k}\right)\cdot\sum_{i=0}^t \frac{n_i}{\left(\sum_{j=0}^i n_j\right)^2} \label{eq:upper-bound-unprocessed}
	\end{align}
\end{corollary}

This bound most closely resembles the results in \cite{gerstgrasser_is_2024, dey_universality_2024}; in fact, if each batch is the same size $n$, we find an upper bound of $\sfrac{\pi^2}{6}\cdot\sfrac{1}{n}$, recovering the additional penalty factor of $\sfrac{\pi^2}{6}$ highlighted therein. We note that this bound increases with $t$, and therefore, in our context, it is perhaps useful only to use the upper bound $\sfrac{1}{n_0}$ obtained by using the order-optimal estimator for just the $0^{\mathrm{th}}$ batch. 

With that said, it is interesting to note that letting the estimate be the empirical distribution of the whole corpus provides a notion of stability `for free'---the sum in the RHS of \eqref{eq:upper-bound-unprocessed} is exactly $\sfrac{1}{n_0}$ plus a (lower) Riemann sum of the function $\sfrac{1}{x^2}$ from $n_0$ to $\sum_{i=0}^t n_i$ with intervals $n_i$, and therefore is always bounded above by a constant smaller than $1$.

In addition, for application in the data replacement workflow, Theorem~\ref{thm:lower-bound} can be modified to obtain the following corollary:

\begin{corollary}[Minimax lower bound with data replacement]\label{cor:lower-bound-replacement}
	Let $\mathbf{Q}_p$ be a distribution on $\Delta_k$ that depends on $p$, and let $\Phat \sim \mathbf{Q}_p$ be an estimate of $p$. Let samples $X_1^{n_1} \sim \tilde{P}({p, \Phat, \alpha})^{n_1}$. The minimax $\ell_2^2$ and $\ell_1$ losses of an estimator $\phat_{1}(X_1^{n_1})$ are lower bounded as
	\begin{align}
		r^{\ell_2^2}_{k, \{\alpha, n_1\}} \geq& \frac{\sfrac{1}{512}}{n_1\alpha^2 + h({\sfrac{1}{4k}})}\\
		r^{\ell_1}_{k, \{\alpha, n_1\}} \geq& \sqrt{\frac{\sfrac{k}{4096}}{n_1\alpha^2 + h({\sfrac{1}{4k}})}}
	\end{align}
	whenever $n_1\alpha^2 + h\left({\sfrac{1}{4k}}\right) \geq k/4$; $h({\epsilon})$ here is defined as $n_1\alpha\sup_p \max_j \mathbf{Q}_p\{\Phat[j] < p[j]-\epsilon\}$.
\end{corollary}

The proofs of Theorem~\ref{thm:lower-bound} and Corollary \ref{cor:lower-bound-replacement} are deferred to Appendix \ref{sec:proofs-lb} and the proofs of Theorem~\ref{thm:upper-bound} and Corollary \ref{cor:upper-bound-unprocessed} are deferred to Appendix \ref{sec:proofs-ub}. Since the bounds for $\ell_1$ loss are proportional to $\sqrt{k}$ times the square root of the $l_2^2$ bounds, we only consider $\ell_2^2$ bounds in the rest of the paper. 

\subsection{Baseline: oracle-assisted minimax loss}
To understand the impact of not knowing the source identity on the minimax loss, it is useful to compare the upper and lower bounds presented above to the loss in the case where the identity of the source of each sample is known in the form of an auxiliary random vector $W := W_i^{n_i} \sim (\mathrm{Bern}(\alpha_i))^{n_i}, i\in \{1, \dots, t\}$ provided to the estimator by an oracle such that, conditioned on whether $W_i[s]$ is $0$ or $1$, the sample $X_i[s]$ is distributed either as $p$ or $\Phat_{i-1}$ respectively.

\begin{lemma}[Oracle-assisted minimax loss]\label{lem:oracle-minimax}
	The minimax loss $r^{\ell_2^2, \mathrm{oracle}}_{t, k}$ of self-consuming estimation with information on source identity is bounded between
	\begin{align*}
		\frac{\sfrac{1}{512}}{\sum\limits_{i\in [0:t]} n_i \alpha_i} \leq r^{\ell_2^2, \mathrm{oracle}}_{t, k} \leq \frac{2}{\sum\limits_{i\in [0:t]} n_i\alpha_i}.
	\end{align*}
\end{lemma}
The proof of this Lemma is deferred to Appendix \ref{sec:proofs-oracle}.

\begin{rem}[Minimax bounds for other losses]
	The bounds in this work have been shown for the $\ell_1$ and $\ell_2^2$ risks, with a focus on the $\ell_2^2$ risk for clarity of exposition. The $\ell_1$ loss is related to other losses used in the model collapse literature via standard inequalities.
	\begin{itemize}
		\item (Total variation and Wasserstein-1 losses): It is well-known that the total variation distance between two distributions is exactly twice their $\ell_1$ distance. In the discrete setting, the Wasserstein $1$-distance is equal to the total variation under the natural metric $\rho(., .) := \openone\{.\neq .\}$.
		\item (KL divergence and cross-entropy losses): Pinsker's inequality relates the total variation to the KL divergence. Additionally, the cross-entropy loss/log-loss $E_p[-\log q(X)]$ used widely in the generative machine learning literature is exactly equal to $D(p\|q) + H(p) $, where $H(.)$ is the entropy of the true distribution.
		\item (Hellinger distance): Le Cam's inequalities \cite[Lemma 2.3]{tsybakov_introduction_2009}
		\begin{align*}
			\frac{1}{2}\mathcal{H}^2(p, q) \leq \|p-q\|_{TV} \leq \mathcal{H}(p, q)\sqrt{1-\frac{\mathcal{H}^2(p, q)}{4}}
		\end{align*} relate the total variation distance and the Hellinger distance $\mathcal{H}(., .)$ respectively.
	\end{itemize}	
	We can therefore directly use the upper and lower bounds provided in this work for the minimax Hellinger distance and the lower bounds for the KL divergence and cross-entropy losses. Using these inequalities, we have lower and upper bounds on the minimax Hellinger distance scaling as the minimax $\ell_1$ distance and its square root respectively; and the lower bound on the minimax KL divergence (and therefore also the minimax excess cross-entropy loss) scaling as the square of the minimax $\ell_1$ loss.
\end{rem}

\section{Discussion}

It is evident from the bounds shown above that if the mixing factor $\alpha_t$ is constant (or even $\Omega(1)$), we can show that the base and oracle-assisted minimax losses are within constant factors of each other; the case where the probability converges to zero is more interesting. This corresponds to cases where the rate of synthetic sampling outpaces the real sampling rate by a considerable margin. Such cases might arise when it is significantly cheaper to sample the synthetic distribution, perhaps also exacerbated by the fact that the estimates get closer to the real distribution with every new batch. The behavior of the bounds in such cases depends on the rates of growth of $\sum_{i=0}^tn_i\alpha_i$ and $\sum_{i=0}^t n_i\alpha_i^2$. The absence of source information at the estimator leads to a penalty factor of roughly $\max(\alpha_i, g_{i-1}(\sfrac{1}{4k}))$ in the `effective sample size' of each batch; we show that this can lead to large significant gaps between the base and oracle-assisted losses. Depending on whether or not the $\alpha_i$'s are comparable to the $g_{i-1}(\sfrac{1}{4k})$'s, there might also be a gap between the guarantees provided by the upper and the lower bounds. In the sequel, we comment on the impact of the error term on the behavior of the bounds and the restrictions on the order-optimal estimator, show cases where differences in the aforementioned rates of growth imply gaps between the base and oracle-assisted losses (suggesting a change in scaling laws), and conclude with a summary of the regimes where it is possible to comment precisely on the minimax error.

\subsection{The error term $h_t({\sfrac{1}{4k}})$}
The multi-stage lower bounds depend on the performance of estimators in previous stages through the probabilities $g_{i}\left(\sfrac{1}{4k}\right)$. Intuitively, the quantities $g_i(\sfrac{1}{4k})$ capture the worst-case deviation probability of \textit{any} component of $\Phat_{i-1}$ from $p$ by more than $\sfrac{1}{4k}$. Since we focus consider cases where $\Phat_t$ converges to $p$, this probability is expected to decay to zero. It is perhaps surprising, then, to also observe that the lower bounds \textit{decrease} as the probabilities $g_i(\sfrac{1}{4k})$ increase---suggesting that it might be possible to improve the performance of future estimators by estimating badly in the present. The source of this apparent improvement is the loosening of the upper bound on the KL divergence in \eqref{eq:kl-upper-bound-expectation} in instances when the previous estimators perform badly. This suggests that it might be easier for an estimator to perform well when the synthetic samples are from a distribution sufficiently removed from the real one; the sequence of estimators in Theorem~\ref{thm:upper-bound} does not share this behavior.

We note that the error term is small compared to the other term in the denominator, $\sum_{i=0}^t n_i\alpha_i^2$, in quite a lot of cases. Using Chebyshev's inequality (as demonstrated in the upcoming examples) leads to an upper bound on each of these probabilities in terms of the minimax error $r_{i-1, k}^{\ell}$; the upper and lower bounds thus match whenever $n_i\alpha_i^2$'s are larger than or comparable to $1/\sum_{j=0}^{i-1}n_{j}\alpha_{j}^2$. We also note that it is necessary to use Chebyshev's inequality---as opposed to stronger concentration results such as Hoeffding-style bounds that rely on the specific properties, such as subgaussianity in the case of the Hoeffding bound, of the estimators---to derive a universal upper bound on these probabilities for \textit{any} sequence of estimators that achieves order-optimal performance.

\subsection{The order-optimal estimator}
The sequence of estimators in Theorem~\ref{thm:upper-bound} do not \textit{always} return distributions; indeed, an assumption on the sample sizes and mixing factors is required to make sure that the outputs stay in $\Delta_k$ at every stage. Each estimate in the sequence is also kept unbiased and uncorrelated from the previous estimate; this helps keep the analysis from becoming almost prohibitively cumbersome. A careful analysis of a more general class of nonlinear estimators might help bridge the gap with the lower bounds, proving the error term necessary, or drop the assumptions required. Alternatively, a tighter analysis bounding the aforementioned divergence might lead to a lower bound that matches the upper bound wherever valid. 

For the rest of this paper, we analyze the lower bounds assuming `good' estimators---estimators with worst-case loss within at most a constant factor of the lower bound---exist, and are chosen at every stage. In quite a few cases, as it turns out, we find that the error term is small as compared to the rest of the denominator, and the analysis is tight if optimal or near-optimal estimators are assumed to have been used. We now show the main results in action via two examples, one where the bounds match and the error term can be ignored, and one where the error term dominates, but still leads to a gap w.r.t.\ the oracle-assisted loss.

\subsection{Demonstrative examples for data accumulation} We first demonstrate choices for the sequence $n_t, \alpha_t, t\geq 0$ where the upper and lower bounds match (up to constant factors) and show a polynomially growing gap from the oracle-assisted minimax loss in the accumulation setting.
\subsubsection{An example with matching bounds} Let $n_i=a(i+1)^{\beta+\gamma}$, and $\alpha_i = (i+1)^{-\gamma}$, where $a, \beta, \gamma$ are positive constants such that $a\geq \sfrac{k}{4}$ and $ \gamma \leq \min (1, \sfrac{1{+}\beta}{2})$. Thus, on average, the number of synthetic samples at stage $t$ is $\Theta(t^{\beta+\gamma})$, while the number of real samples is $\Theta(t^\beta)$. The assumption on $a$ ensures that the lower bound (Theorem~\ref{thm:lower-bound}) holds for every $t\geq 1$, and the assumptions on $\gamma$ ensure that the upper bound (Theorem~\ref{thm:upper-bound}) is valid and the error term $h_{\sfrac{1}{4k}}(t)$ is, in the worst case, comparable to $\sum_{i=0}^tn_i\alpha_i^2$ for each $t$. Additionally, for the base setting, assume that the estimators in Theorem~\ref{thm:upper-bound} are used at each stage.
\begin{claim}\label{ex:matching}
	For the parameters described above, the minimax loss $r^{\ell_2^2}_{t, k}$ is $\Theta(t^{-(\beta - \gamma+1)})$, and the oracle-assisted minimax loss $r^{\ell_2^2, \mathrm{oracle}}_{t, k}$ is $\Theta(t^{-(\beta+1)})$.
\end{claim}
\begin{IEEEproof}
	First, note that $\sum_{i=0}^{t}n_i\alpha_{i} = \sum_{i=0}^t a(i+1)^\beta = \Theta \left((t+1)^{\beta+1}\right)$. Using Lemma~\ref{lem:oracle-minimax}, \begin{align*}
		\frac{c_1(1+\beta)}{a(t+1)^{1+\beta}} \leq r^{\ell_2^2, \mathrm{oracle}}_{t, k} \leq \frac{c_2(1+\beta)}{a(t+1)^{1+\beta}} .
	\end{align*} In contrast, for the base case, using Theorem~\ref{thm:upper-bound}, since $\sum_{i=0}^{t}n_i\alpha_i^2 = a\sum_{i=0}^{t} (i+1)^{\beta-\gamma} \geq a\frac{(t+1)^{\beta-\gamma+1}}{\beta-\gamma+1}$,
	\begin{align*}
		r^{\ell_2^2}_{t, k} \leq \frac{\left(1-\frac{1}{k}\right)(\beta {-} \gamma {+}1)}{a (t+1)^{\beta {-} \gamma {+}1}}.
	\end{align*}
	Since the variance of each component $j$ of each estimate $\Phat_{i-1}$ is bounded above as $\var\left(\Phat_{i-1}[j]\right) \leq E\left[\ell_2^2\left(\Phat_{i-1}, p\right)\right]\leq r_{i-1, k}^{\ell_2^2}$ and $\Phat_{i-1}$ is unbiased for every $i$, using Chebyshev's inequality on $\Pbar_{i-1,({p})}\left\{\Phat_{i-1}[j]< p[j]-\sfrac{1}{4k}\right\}$,
	\begin{align*}
		h_t(\sfrac{1}{4k}) =& \sum_{i=1}^t n_i\alpha_i g_i(\sfrac{1}{4k}) \\
		=& \sum_{i=1}^t n_i\alpha_i \sup_p \max_j \Pbar_{i-1,({p})}\left\{\Phat_{i-1}[j]< p[j]{-}\frac{1}{4k}\right\} \\
		\leq & \sum_{i=1}^t a(i+1)^\beta \cdot \frac{\var\left(\Phat_{i-1}\right)}{\left(\sfrac{1}{4k}\right)^2}\\
		 \leq & \sum_{i=1}^t a(i+1)^\beta \cdot 16k^2 \cdot r_{i-1, k}^{\ell_2^2} \\
		\leq & \sum_{i=1}^t a(i+1)^\beta \cdot 16k^2 \cdot \frac{\left(1-\frac{1}{k}\right)(\beta {-} \gamma {+}1)}{a \cdot i^{\beta {-} \gamma {+}1}}\\
		 \leq & 2^\beta\cdot16k^2\cdot(\beta{-}\gamma{+}1)\sum_{i=1}^ti^{\gamma-1} \numberthis\label{eq:chebyshev-reference}\\
		\leq & 2^\beta\cdot16k^2\cdot (\beta {-} \gamma {+} 1)\frac{t^\gamma}{\gamma}.
	\end{align*}
	Plugging this into Theorem~\ref{thm:lower-bound}, since $\gamma \leq \beta {-} \gamma {+}1$ by assumption, there exists $c_3 \in \mathbb{R}_+$ such that
	\begin{align*}
		r^{\ell_2^2}_{t, k} \geq \frac{c_3}{(t+1)^{\beta - \gamma+1}}.
	\end{align*}
\end{IEEEproof}
This inductively shows that the upper and lower bounds match in terms of order w.r.t.\ the stage $t$: for each stage $t$, the upper bound is within a constant factor of the lower bound, provided that order-optimal estimators are chosen for stages $0$ through $t{-}1$. Moreover, there is a polynomially growing gap between the base and oracle-assisted minimax losses:
\begin{align}
	\frac{r^{\ell_2^2}_{t, k}}{r^{\ell_2^2, \mathrm{oracle}}_{t, k}}= \Theta(t^\gamma). \label{eq:rate-difference-polynomial}
\end{align}
The general regime where the bounds match is given in Proposition~\ref{prop:matching}.

\subsubsection{An example where the upper and lower bound do not match} Consider the previous example but with $1 > \gamma > \frac{1{+}\beta}{2}$ instead. The upper bound on $\gamma$ ensures that the upper bound from Theorem~\ref{thm:upper-bound} is still valid, but the error term will now dominate the analysis of the lower bound (Theorem~\ref{thm:lower-bound}). The effect of the error term growing large is seen in the following Proposition.

\begin{claim} \label{ex:non-matching}
	Consider the self-consuming distribution estimation problem with the parameters given above. For $t\geq 1$, assume that, for stages $ i\in [0:t-1]$, there exists a sequence of estimators for which the worst-case expected loss is smaller than $M\cdot\sfrac{1}{(i+1)^{\frac{\beta{+}1}{2}}}$ for a positive constant $M$. If these estimators are used in the first $t$ stages, then there exists a positive constant $m$ independent of $t$ such that the minimax loss $r^{\ell_2^2}_{t, k}$ at stage $t$ is larger than $m\cdot\sfrac{1}{(t+1)^{\frac{\beta{+}1}{2}}}$. The worst-case expected loss for the estimators in Theorem~\ref{thm:upper-bound} remains $\Oc(\sfrac{1}{(t+1)^{\beta+1-\gamma}})$.
\end{claim}
\begin{IEEEproof}
	Let the worst-case expected loss of the sequence of estimators at stage $t$ be denoted by $\sigmab_t^2$, with $\sigmab_{-1}^2 = 0$. Then for $t\geq1$, using Theorem~\ref{thm:lower-bound} and Chebyshev's inequality (applied in the same way as \eqref{eq:chebyshev-reference})
	\begin{align}
		\sigmab_t^2 \geq & \frac{1}{512a\sum_{i=0}^t (i+1)^{\beta-\gamma} + \frac{(i+1)^\beta k^2\sigmab_{i-1}^2}{32}}.\label{eq:polynomial-recursion}
	\end{align}
	Now, by assumption, for $i\in [0:t-1]$, \begin{align*}
		\sigmab_i^2 \leq&  \frac{M}{(i+1)^{\sfrac{(\beta{+}1)}{2}}} \numberthis\label{eq:induction-assumption}\\
		\implies \sum_{i=1}^{t} (i+1)^\beta\sigmab_{i-1}^2 \leq M\sum_{i=1}^{t} \left(1+\frac{1}{i}\right)^\beta i^{\frac{\beta-1}{2}}
		\leq & 2^\beta \cdot M \sum_{i=1}^{t} i^{\frac{\beta-1}{2}}
		\leq  2^\beta \cdot M \cdot \frac{2}{\beta+1}\cdot (t+1)^{\frac{\beta+1}{2}}.
	\end{align*} Also, since $\beta-\gamma+1 < \sfrac{\beta{+}1}{2}$ by assumption, $\sum_{i=0}^t(i+1)^{\beta-\gamma} = o\left(\left(t+1\right)^{\frac{\beta+1}{2}}\right)$. Substituting these into \eqref{eq:polynomial-recursion}, \begin{align*}
		\sigmab_t^2 \geq & \frac{1}{512a\cdot \frac{k^2}{32}\cdot 2^\beta \cdot M \cdot \frac{2}{\beta+1}\cdot (t+1)^{\frac{\beta+1}{2}} + o\left(\left(t+1\right)^{\frac{\beta+1}{2}}\right)},
	\end{align*}setting $m=512a\cdot\frac{k^2}{32}\cdot 2^\beta \cdot M \cdot \frac{2}{\beta+1} + \epsilon$, where $\epsilon$ accounts for the $o(\cdot)$ term, completes the proof.
	
	The upper bound on the worst-case loss of the estimators in Theorem~\ref{thm:upper-bound} follows unchanged from Example \ref{ex:matching}.
\end{IEEEproof}

\begin{figure}[t]
	\centering
	\includegraphics[width=0.4\linewidth]{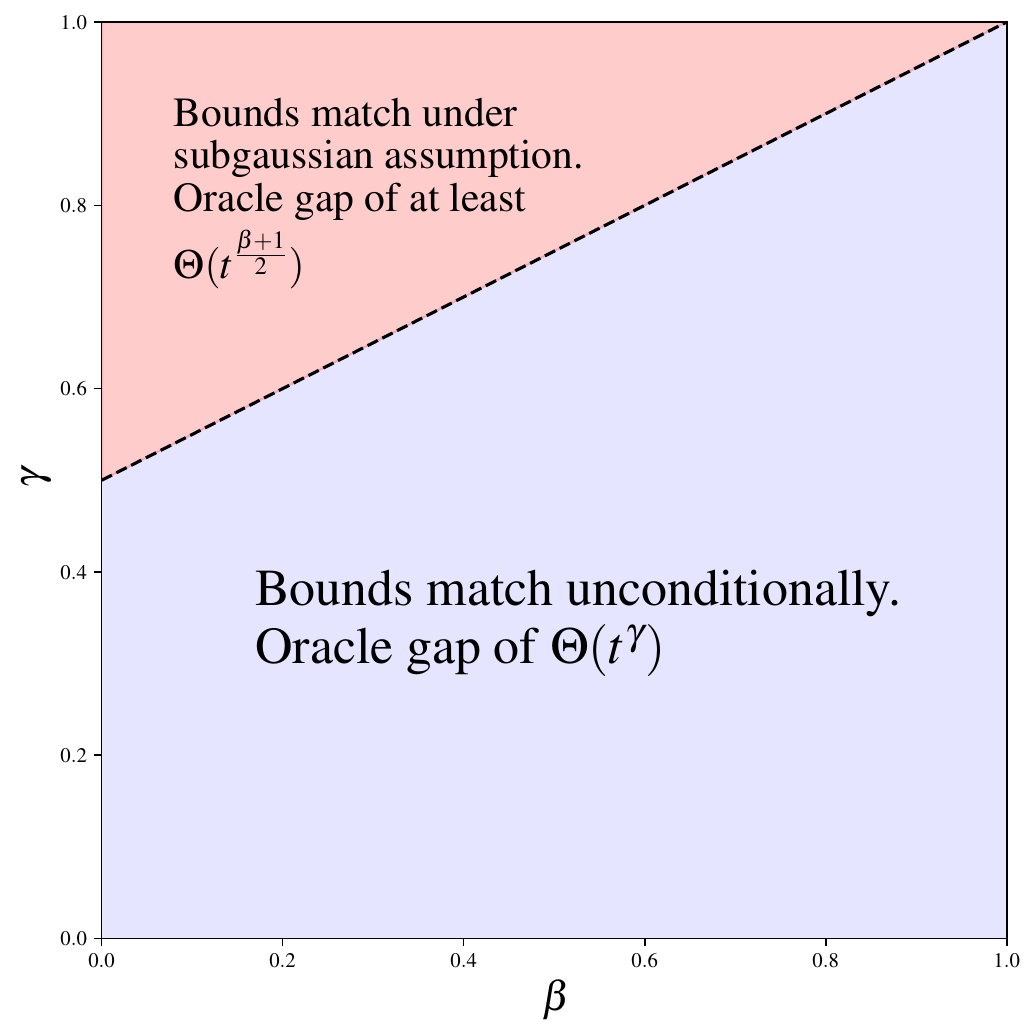}
	\caption{Regimes of minimax loss in the data accumulation workflow with $\Theta(t^{\beta})$ real and $\Theta(t^{\beta+\gamma})$ synthetic samples at stage $t$; gaps indicated are w.r.t. the oracle-assisted loss}
	\label{fig:regimesconvergence}
\end{figure}

Comparing the lower bound against the oracle-assisted minimax loss shows that there is a gap of at least $\Theta(t^{\sfrac{\beta{+}1}{2}})$. The gap with the upper bound remains $\Theta(t^\gamma)$. Since the error term $h_t(\sfrac{1}{4k})$ dominates the analysis of the lower bound, the impact of a much faster decaying sequence of $\alpha_t$'s on the lower bound is limited. This is not to say that the impact of the penalty is small: the effective net sample size still becomes $\Oc(\sqrt{t^{\beta+1}})$ as opposed to $\Theta(t^{\beta+1})$ with oracle information. The two regimes for $\beta, \gamma\in [0, 1]^2$ are depicted in Figure \ref{fig:regimesconvergence}.

We once again note that stronger upper bounds, for example, of the form $h_t(\sfrac{1}{4k}) \leq \exp(-\lambda_0 t)$, can be assumed by restricting the class of estimators appropriately; this would then make the estimators in Theorem~\ref{thm:upper-bound} order optimal for every choice of parameters that satisfies the hypothesis condition. Bounds using the Chebyshev inequality remain the tightest without such assumptions.

\subsection{General regimes for data accumulation}
We first show that when $\alpha_t$ is $\Omega(1)$, the minimax error coincides with the oracle-assisted minimax error up to constant factors.
\begin{proposition}\label{prop:basic}
	When $\alpha_t\geq l>0$ for all $t\geq 0$, there exist constants $m$ and $M$ independent of $t$ such that
	\begin{align*}
		\frac{m}{\sum_{i=0}^tn_i\alpha_{i}} \leq r^{\ell_2^2}_{t, k} \leq \frac{M}{\sum_{i=0}^tn_i\alpha_{i}}
	\end{align*}
\end{proposition}
\begin{IEEEproof}
	It is easy to see that since $l\leq \alpha_t\leq1$, $l\sum_{i=0}^tn_i\alpha_i\leq \sum_{i=0}^t n_i\alpha_i^2\leq \sum_{i=0}^{t}n_i\alpha_i$. The bounds follow directly from Theorems \ref{thm:lower-bound} and \ref{thm:upper-bound}.
\end{IEEEproof}

We now describe the more interesting regime where the upper and lower bounds on the minimax loss match in the following Proposition. It asserts that when $\alpha_t$ vanishes with $t$, the bounds match if the $\ell_2^2$ loss vanishes faster.
\begin{proposition}\label{prop:matching}
	When $\alpha_t\downarrow0$, the ratio between the upper and the lower bounds on the minimax $\ell_2^2$ loss (Theorems \ref{thm:upper-bound} and \ref{thm:lower-bound} respectively) is bounded above by a constant independent of the stage $t$ if $\alpha_t = \Omega\left(\sfrac{1}{\sum_{i=0}^{t-1} n_i\alpha_i^2}\right)$. Moreover, if $\sfrac{1}{\sum_{i=0}^{t-1} n_i\alpha_i^2} = o(\alpha_t) $, there exists $t_0\geq 0$ such that the ratio is also bounded above by a constant independent of $k$ for stages $t\geq t_0$.
\end{proposition} 
\begin{IEEEproof}[Outline of proof]
	Using Chebyshev's inequality as shown above, along with the assumption on $\alpha_t$, ensures that $g_t(\sfrac{1}{4k})$ is smaller than $16k^2\cdot m \alpha_t$ for a positive constant $m$, and therefore $n_i\alpha_i g_i(\sfrac{1}{4k})$ is smaller than $16k^2 \cdot m n_i\alpha_i^2$; the lower bound is then
	\begin{align}
		r^{l_2^2}_{t, k} \geq \frac{\sfrac{1}{512}}{\left(1+16k^2m\right)\sum_{i=0}^t n_i\alpha_i^2}, \label{eq:prop-matching-lb}
	\end{align}completing the first part of the proof. If $\sfrac{1}{\sum_{i=0}^{t-1} n_i\alpha_i^2} = o(\alpha_t) $, $\sum_{i=0}^{t} n_i\alpha_i^2$ also diverges (since $\alpha_t\leq 1$), and one can find $t_1$ such that $g_t(\sfrac{1}{4k})$ is smaller than $\epsilon$---independent of $k$---for $t\geq t_1$. The denominator of \eqref{eq:prop-matching-lb} can then be split into $(1+16k^2m)\sum_{i=0}^{t_1-1}n_i\alpha_{i}^2$ and $(1+\epsilon)\sum_{i=t_1}^{t}n_i\alpha_{i}^2$, the latter of which will dominate the former as $t$ grows large.
\end{IEEEproof}

If $\alpha_t$ decays too quickly, the error term starts dominating the denominator of the lower bound. In spite of this, we have the following inductive lower bound on the minimax loss:
\begin{proposition}\label{prop:non-matching}
	Consider the self-consuming distribution estimation problem with parameters $n_t, \alpha_t$ such that \begin{enumerate}
		\item $\alpha_t\downarrow0$,
		\item $\alpha_t = \Oc(1/\sum_{i=0}^{t-1} n_i\alpha_i^2)$, 
		\item $n_t\alpha_t = \Oc(\sum_{i=0}^{t-1} n_i\alpha_i)$
		\item $\sum_{i=0}^t n_i\alpha_i \uparrow \infty$.
	\end{enumerate} Assume that there exist estimators $\phat_{i}(\cdot)$ such that the worst case $\ell_2^2$ loss is bounded as\begin{align*}
		\sup_pE[\ell_2^2(\Phat_{i}, p)] \leq \frac{M}{\sqrt{\sum_{j=0}^i n_j\alpha_j}}
	\end{align*} for each $i$ in $[0:t-1]$, for some $M>0$. Then there exists a constant $m>0$ such that the minimax loss at stage $t$ is bounded below as
	\begin{align*}
		r^{\ell_2^2}_{t, k} \geq \frac{m}{\sqrt{\sum_{i=0}^t n_i\alpha_i}}
	\end{align*}when the aforementioned estimators are selected for stages $0$ to $t-1$.
\end{proposition}
The proof of this Proposition is deferred to Appendix \ref{app:non-matching}.
The third condition here is satisfied by all sequences for which $n_t\alpha_t$ grows exponentially or slower (as can be seen by the Discrete Gronwall inequality \cite{clark_short_1987}). This proposition solidifies the intuition for the lower bound when the mixing probabilities $\alpha_t$ decay too quickly: the effective sample size is proportional to the square root of the sample size when in the oracle-assisted setting.

\subsection{Data replacement} 
In this section, we demonstrate the application of Corollary \ref{cor:lower-bound-replacement} in the data replacement setting where, at every stage $t$, the estimator only has access to samples generated in the current stage $X_t^n$. Note that $\alpha$ and $n$ do not change with $t$ in this setting, except at stage $0$ where all $n$ samples are real. This setting has been explored in depth in the literature e.g.\ \cite{alemohammad2024selfconsuming, bertrand2024on, seddik2024how, shumailov_ai_2024, dohmatob_model_2024, fu_towards_2024, dohmatob2025strong, suresh2025rate}. 

Let the worst-case expected $\ell_2^2$ loss at stage $t$ be denoted as $\sigmab_t^2$, and define $M := \sup_{t\geq 0} \sigmab_t^2$. Repeatedly applying Corollary \ref{cor:lower-bound-replacement}, for every $t\geq 0$,
\begin{align}
	\sigmab_{t+1}^2 \geq &\ \frac{1}{512\left(n\alpha^2 + n\alpha \sup_{p'}\max_j\Pbar_{t,({p'})}\left\{\Phat_{t}[j]-p'[j] < \sfrac{1}{4k} \right\}\right)}.\label{eq:data-replacement-lb}
\end{align}
Once again using Chebyshev's inequality,
\begin{align*}
	\Pbar_{t,({p'})}\left\{\Phat_{t}[j]-p'[j] < \sfrac{1}{4k} \right\} \leq& \frac{E_{\Pbar_{t,({p'})}}\left[\left(\Phat_{t}[j]-p'[j]\right)^2\right]}{\sfrac{1}{(4k)^2}}
	 \leq  16k^2\cdot\sigmab_t^2 \leq 16k^2M.
\end{align*}Substituting this into \eqref{eq:data-replacement-lb}, the smallest worst-case error has the lower bound
\begin{align*}
	m:= \inf_{t \geq 1} \sigmab_t^2 \geq \frac{1}{512\left(n\alpha^2 + M \cdot 16n\alpha k^2\right)}.
\end{align*}
When $k$ is fixed, we find that since $M\leq \sqrt{2}$ trivially, $m \geq \sfrac{c}{n\alpha}$ for some $c\geq 0$. This is especially stark in contrast to the oracle-assisted infimum, which, with probability $1$, is proportional to $\sfrac{1}{n}$, corresponding to the case where all samples are real. Thus, in the data replacement setting, the `best-case' minimax expected loss is of the same order as the `average' oracle-assisted loss. This result can also be interpreted as an `uncertainty principle' of recursive estimation in the replacement workflow: the above inequality directly leads to a lower bound on $m\cdot M$.

\section{Conclusion and Future Work}
In this work, we propose and study the minimax discrete distribution estimation problem in self-consuming loops, inspired by recent interest in the empirical machine learning literature. We find minimax bounds on the expected loss, and show the behavior of these bounds in various regimes. We discuss the implications of these results on model collapse considerations, particularly in observing changes in scaling laws.

It should, however, be noted that these results do not \textit{directly} apply to practical settings. This is in part because we require in our lower bounds that the effective number of samples is comparable to the alphabet size of $X$, which might be very large in practical settings. The class of distributions under consideration in these settings might also be much smaller than the class of distributions considered in the minimax approach. We conjecture that the behavior of the minimax rates in the restricted setting (referred to in the distribution estimation literature as `local' minimax rates) on the number of samples $n_t$ and the mixing factors $\alpha_t$ remains the same, and the restriction on the space simply changes the dependence on the dimension.

\allowdisplaybreaks

\appendices
\section{Proofs of the minimax lower bounds}\label{sec:proofs-lb}
We assume that the alphabet size $k$ is even. The worst-case loss for odd $k$ is lower bounded by the worst-case loss for $k-1$, since the supremum is over a smaller set of distributions.

The lower bound argument uses Assouad's method with the following construction, widely known in the distribution estimation literature e.g. as described in \cite{paninski_coincidence-based_2008, tsybakov_introduction_2009, yu_assouad_1997}.
Consider the set of vectors $\Vc = \{+1, -1\}^{k/2}$. With each vector $v\in\Vc$, associate the distribution
\begin{align*}
	p_v := p_u + \frac{\delta}{k}\begin{bmatrix}
		v\\-v
	\end{bmatrix}
\end{align*} where $p_u$ is the uniform distribution. Let $\Pc_k:= \{p_v:v\in \Vc\}$.

Assouad's method leads to the lower bound described in the following Lemma. The full proof is provided in Appendix \ref{app:proofs-lemmas} for completeness.
\begin{lemma}\label{lem:lower-bound-partial}
	The minimax estimation risk described in Definition \ref{def:minimax-loss} is lower bounded as
	\begin{align}
		r^{\ell_2^2}_{t, k} \geq \frac{\delta^2}{4k}\left(1 {-} \sqrt{\frac{2}{k}\sum_{j=1}^{k/2} \max_{\substack{v\in\Vc:\\v_j=1}} D\Big(\Pbar_{t, (p_v)} \Big\| \Pbar_{t, (p_{v-2e_j})}\Big)}\right) \label{eq:lower-bound-l2-kl-main}\\
		r^{\ell_1}_{t, k} \geq \frac{\delta}{4}\left(1 - \sqrt{\frac{2}{k}\sum_{j=1}^{k/2} \max_{\substack{v\in\Vc:\\v_j=1}} D\Big(\Pbar_{t, (p_v)} \Big\| \Pbar_{t, (p_{v-2e_j})}\Big)}\right) \label{eq:lower-bound-l1-kl-main}
	\end{align}
\end{lemma}

The following elementary Lemma will be useful in the proof of Theorem~\ref{thm:lower-bound}:
\begin{lemma}\label{lem:exp-upper-bound}
	Let $Z \sim Q$ be a non-negative random variable and $a$, $b$, and $c$ be some positive constants. Then
	\begin{align*}
		E_Q\left[\frac{1}{a + bZ}\right] \leq \frac{Q\{Z< c\}}{a} + \frac{1}{a + bc}.
	\end{align*}
\end{lemma}

\begin{IEEEproof}[Proof of Theorem~\ref{thm:lower-bound}]
	Using the chain rule for KL divergences, we find the upper bound \eqref{eq:kl-upper-bound-expectation}, whenever $j\leq \sfrac{k}{2}, v[j] = +1$, as follows:
	\begin{align*}
		D\Big(\Pbar_{t, (p_v)}  \Big\| \Pbar_{t, (p_{v-2e_j})}\Big)		=& n_0 D\left(p_v\|p_{v-2e_j}\right) + \sum_{i=1}^{t} D\left(\left(\tilde{P}({p_v, \Phat_i, \alpha_{i}})\right)^{n_{i}}  \middle\| \left(\tilde{P}({p_{v-2e_j}, \Phat_i, \alpha_{i}})\right)^{n_{i}} \middle| \Pbar_{i-1, (p_v)}  \right) \\
		=& n_0 D\left(p_v\|p_{v-2e_j}\right) \\
		&+ \sum_{i=1}^{t} n_{i}E_{\Pbar_{i-1, (p_v)}}\left[\sum_{j'=1}^k\left(\alpha_i p_v[j'] + (1-\alpha_i)\Phat_{i-1}[j']\right)\log\frac{\alpha_i p_v[j'] + (1-\alpha_i)\Phat_{i-1}[j']}{\alpha_i p_{v-2e_j}[j'] + (1-\alpha_i)\Phat_{i-1}[j']}\right]\\
		\overset{(a)}{\leq}& n_0 \sum_{j'=1}^k p_v[j']\frac{p_v[j'] - p_{v-2ej}[j']}{p_{v-2ej}[j']} \\
		& + \sum_{i=1}^{t} n_i E_{\Pbar_{i-1, (p_v)}}\left[\sum_{j'=1}^k\left(\alpha_i p_v[j'] + (1-\alpha_i)\Phat_{i-1}[j']\right)\frac{\alpha_i \left(p_v[j'] - p_{v-2ej}[j']\right) }{\alpha_i p_{v-2e_j}[j'] + (1-\alpha_i)\Phat_{i-1}[j']}\right]\\
		\overset{(b)}{=}& n_0 \frac{2\delta}{k}\left(\frac{p_v[j]}{p_{v-2e_j}[j]} - \frac{p_v\left[j+\sfrac{k}{2}\right]}{p_{v-2e_j}\left[j+\sfrac{k}{2}\right]} \right)\\
		&+\sum_{i=1}^t n_i\left(\frac{2\alpha_i\delta}{k}\right)		E_{\Pbar_{i-1, (p_v)}}\Bigg[\frac{\alpha_i p_{v}[j] + (1-\alpha_i)\Phat_{i-1}[j]}{\alpha_i p_{v-2e_j}[j] + (1-\alpha_i)\Phat_{i-1}[j]} - 1 \\
		& \hspace{4.4cm}
		+1-\frac{\alpha_i p_{v}\left[j+\frac{k}{2}\right] + (1-\alpha_i)\Phat_{i-1}\left[j+\frac{k}{2}\right]}{\alpha_i p_{v-2e_j}\left[j+\frac{k}{2}\right] + (1-\alpha_i)\Phat_{i-1}\left[j+\frac{k}{2}\right]}\Bigg]\\
		\overset{(c)}{=}& n_0 \frac{8\delta^2}{k(1-\delta^2)} +\sum_{i=1}^t n_i\left(\frac{2\alpha_i\delta}{k}\right)^2	E_{\Pbar_{i-1, (p_v)}}\Bigg[\frac{1}{\frac{\alpha_i(1-\delta)}{k} + (1-\alpha_i)\Phat_{i-1}[j]}\\
		&\hspace{6.5cm} + \frac{1}{\frac{\alpha_i(1+\delta)}{k} + (1-\alpha_i)\Phat_{i-1}\left[j+\sfrac{k}{2}\right]}\Bigg],\\
		\overset{(d)}{\leq}& n_0 \frac{8\delta^2}{k(1-\delta^2)}+\sum_{i=1}^t n_i \left(\frac{2\alpha_i\delta}{k}\right)^2	\Bigg(\frac{1}{\alpha_{i}\frac{1-\delta}{k} + (1-\alpha_i)\left(\frac{1+\delta}{k} - \frac{1}{4k}\right)}+ \frac{1}{\alpha_{i}\frac{1+\delta}{k} + (1-\alpha_i)\left(\frac{1-\delta}{k} - \frac{1}{4k}\right)}\\
		&\hspace{5cm} + \frac{\Pbar_{i-1, (p_v)}\left\{\Phat_{i-1}[j] < \frac{1+\delta}{k} - \frac{1}{4k}\right\}}{\alpha_{i}\frac{1-\delta}{k}} \\
		&\hspace{5cm} + \frac{\Pbar_{i-1, (p_v)}\left\{\Phat_{i-1}\left[j+\frac{k}{2}\right] < \frac{1-\delta}{k} - \frac{1}{4k}\right\}}{\alpha_{i}\frac{1+\delta}{k}} \Bigg)\\
		\overset{(e)}{\leq}& n_0 \frac{8\delta^2}{k(1-\delta^2)}+\sum_{i=1}^t n_i \frac{(2\alpha_i\delta)^2}{k}\Bigg(\frac{1}{\alpha_{i}({1-\delta}) + (1-\alpha_i)\left(\frac{3}{4}+\delta\right)} + \frac{g_i(\sfrac{1}{4k})}{\alpha_{i}({1-\delta})}\\
		&\hspace{5cm} +\frac{1}{\alpha_{i}(1+\delta) + (1-\alpha_i)\left(\frac{3}{4}-\delta\right)} + \frac{g_i(\sfrac{1}{4k})}{\alpha_{i}(1+\delta)} \Bigg).\numberthis\label{eq:kl-upper-bound-expectation}
	\end{align*} Here,	$(a)$ is a consequence of $\log x \leq x-1$, $(b)$ and $(c)$ follow from the construction of $p_v$, $(d)$ follows from Lemma~\ref{lem:exp-upper-bound}, and $(e)$ follows from Definition \ref{def:error-term}.
	Now, note that
	\begin{align*}
		\frac{1}{\alpha_{i}(1{-}\delta) + (1{-}\alpha_i)\left(\frac{3}{4}{+}\delta\right)} + \frac{1}{\alpha_{i}(1{+}\delta) + (1{-}\alpha_i)\left(\frac{3}{4}{-}\delta\right)} = \frac{2a_i}{a_i^2 - b_i^2\delta^2}\leq \frac{2}{\frac{9}{16} - \delta^2},
	\end{align*}where $a_i := \sfrac{3}{4}(1-\alpha_{i}) + \alpha_{i} $ and $b_i := 1-2\alpha_i$. The inequality holds since $a_i \in [\sfrac{3}{4}, 1]$ and $b_i\in [-1, 1]$. Thus, we finally get
	\begin{align*}
		D\Big(\Pbar_{t, (p_v)}  \Big\| \Pbar_{t, (p_{v-2e_i})}\Big) \leq  \frac{8\delta^2 \sum_{i=0}^t n_i\alpha_i^2+n_i\alpha_{i}g_{i}(\sfrac{1}{4k})}{k\left(\frac{9}{16} - \delta^2\right)}.\numberthis\label{eq:upper-bound-kl}
	\end{align*}
	Assuming $\sum_{i=1}^t n_i\alpha_i (\alpha_i + g_{i}(\sfrac{1}{4k})) \geq \sfrac{k}{4}$ and choosing \begin{align}
		\delta^2 = \frac{k}{64\sum_{i=0}^t n_i\alpha_i (\alpha_i + g_i(\sfrac{1}{4k}))} <1\label{eq:lb-delta-def}
	\end{align} ensures that $D\left( \Pbar_{t, (p_v)} \middle\| \Pbar_{t, (p_{v-2e_j})}\right) \leq \sfrac{1}{2}$. 
	Invoking Lemma~\ref{lem:lower-bound-partial} and substituting \eqref{eq:lb-delta-def} and \eqref{eq:upper-bound-kl} into \eqref{eq:lower-bound-l2-kl-main} and \eqref{eq:lower-bound-l1-kl-main} concludes the proof.
\end{IEEEproof}

\begin{IEEEproof}[Proof of Corollary \ref{cor:lower-bound-replacement}]
	A slightly modified version of Lemma~\ref{lem:lower-bound-partial} leads to the same lower bounds as \eqref{eq:lower-bound-l2-kl-main}, \eqref{eq:lower-bound-l1-kl-main}, but with the KL divergence term $D\Big(E_{Q_{p_v}}\left[\tilde{P}({p_v, \Phat, \alpha})\right]\Big\| \ E_{Q_{p_v}}\left[\tilde{P}({p_{v-2e_j}, \Phat, \alpha})\right]\Big)$ instead, which has the upper bound
	\begin{align*}
		D\Big(E_{Q_{p_v}}[\tilde{P}({p_v, \Phat, \alpha})]\Big\| \ E_{Q_{p_v}}[\tilde{P}({p_{v-2e_j}, \Phat, \alpha})]\Big)
		 \leq E_{Q_{p_v}}\left[D\left(\tilde{P}({p_v, \Phat, \alpha})\Big\| \tilde{P}({p_{v-2e_i}, \Phat, \alpha})\right)\right].
	\end{align*}The rest of the proof follows the same steps as the proof of Theorem~\ref{thm:lower-bound}.
\end{IEEEproof}

\section{Proofs of the minimax upper bounds}\label{sec:proofs-ub}
In this section, we derive upper bounds on the minimax expected loss at the $t^{\mathrm{th}}$ stage. 

\subsection{The order optimal estimator}
We first describe the performance of a sequence of estimators that are order-optimal with respect to $n_t$ and $\alpha_t$ under mild conditions; finding optimal estimator sequences in other regimes remains an interesting open problem.

Let $\phat_{\mathrm{emp}}(\cdot)$ be the empirical estimator of a given batch of samples. We have the following Lemma:
\begin{lemma}\label{lem:upper-bound-uncombined}
	Let $\Phat_{0}$ be an unbiased estimate of $p$ with a variance of $\eta\cdot p[j](1-p[j]) \geq 0$ for each component $j$. Let samples $X_1^{n_1} \sim \tilde{P}({p, \Phat_0, \alpha})^{n_1}$ as described in Definition \ref{def:self-consuming-seq}. Then the estimator
	\begin{align*}
		\Phat_{1, (\mathrm{cond})} =\phat_{1, (\mathrm{cond})}(X_1^{n_1}) := \frac{1}{\alpha}\left(\phat_{\mathrm{emp}}(X_1^{n_1}) - (1-\alpha)\Phat_0\right) \numberthis\label{eq:conditionally-unbiased-estimator}
	\end{align*}is an unbiased estimator of $p$ and satisfies $E\left[\Phat_0[j]\cdot\Phat_{1, (\mathrm{cond})}[j]\right] = p[j]^2$ for every $j\in [1:k]$.
	Additionally, each component $j$ has variance\begin{align*}
		E\left[\left(\Phat_{1, (\mathrm{cond})}[j] {-} p[j]\right)^2\right] = \frac{p[j](1{-}p[j])}{n_1\alpha^2}\left(1 {-} (1{-}\alpha)^2\eta\right). \numberthis\label{eq:self-consuming-upper-bound}
	\end{align*}
\end{lemma}The proof of this Lemma is deferred to Appendix \ref{app:proofs-lemmas}.

Using elementary calculations, we also have the following Lemma:
\begin{lemma}\label{lem:upper-bound-gen-combination}
	Let $Y_0, Y_1$ be unbiased estimates of a scalar $\theta$ such that $E[Y_0Y_1] = \theta^2$ and $E[(Y_j- \theta)^2] = \sigma_j^2$ for $j=0, 1$. For $a_0, a_1\in \mathbb{R}$,
	\begin{align*}
		E[(a_0Y_0 {+} a_1 Y_1 {-} \theta)^2] = a_0^2\sigma_0^2 {+} a_1^2\sigma_1^2 {+} ((a_0{+}a_1) - 1)^2\theta^2.
	\end{align*}
\end{lemma}
Combining Lemmas \ref{lem:upper-bound-uncombined} and \ref{lem:upper-bound-gen-combination}, we arrive at the following intermediate result:
\begin{lemma}\label{lem:upper-bound-one-step}
	Let $\Phat_{0} = \phat_{0}(X_0^{n_0})$ be an unbiased estimate of $p$ with variance $\eta\cdot\left(\sum_j p[j](1-p[j])\right)$ for each component $j$. For $X_1^{n_1}\sim \tilde{P}({p, \Phat_0[j], \alpha})^{n_1}$, there exists an unbiased estimator $\phat_{1}(\cdot)$ such that for $\Phat_1 := \phat_{1}(X_0^{n_0}, X_1^{n_1})$, for each component $j$,
	\begin{align}
		E_{P_{X_0^{n_0}, X_1^{n_1}}}\left[\left(\Phat_1[j]- p[j]\right)^2\right] \leq \frac{p[j](1-p[j])}{\frac{1}{\eta} + n_1\alpha^2}.
	\end{align}
	Moreover, the estimate $\Phat_{1}$ thus obtained is a valid distribution almost surely if $\alpha - \alpha^2 \leq \frac{1}{\eta n_1}$.
\end{lemma}
\begin{IEEEproof}
	Let $\Phat_{1} = a_0\Phat_0 + a_1\Phat_{1, (\mathrm{cond})}$ with \begin{align*}
		a_0= \frac{\sfrac{1}{\eta}}{\sfrac{1}{\eta} + \sfrac{1}{(n_1\alpha^2)}}  \mbox { and }		a_1= 1-a_0.
	\end{align*} The resulting estimator can then be expressed as
	\begin{align}
		\phat_1(X_0^{n_0},& X_1^{n_1})= \phat_0(X_0^{n_0}) + \frac{n_1\alpha}{\frac{1}{\eta} + n_1\alpha^2}\left(\phat_{\mathrm{emp}}(X_1^{n_1}) - \phat_0(X_0^{n_0})\right).\label{eq:upper-bound-estimator}
	\end{align}Now, $\alpha-\alpha^2 \leq \frac{1}{\eta n_1}$ ensures that the RHS of \eqref{eq:upper-bound-estimator} is a convex combination of two distributions in $\Delta_k$, thus ensuring that $\Phat^{(1)}$ is a valid distribution almost surely. Using Lemmas \ref{lem:upper-bound-uncombined} and \ref{lem:upper-bound-gen-combination} leads to the desired upper bound on the componentwise variances.
\end{IEEEproof}

\begin{IEEEproof}[Proof of Theorem~\ref{thm:upper-bound}]
	If $\phat_0$ is the empirical estimator, it is well known that $E[\ell_2^2(\Phat_0[j], p[j])] = \sfrac{p[j] (1-p[j])}{n_0}$. Proceeding inductively and repeatedly applying Lemma~\ref{lem:upper-bound-one-step}, we find that the estimated distribution at each stage $t$ has mean $p$, and thus the expected $\ell_2^2$ loss is precisely the sum of the componentwise variances, each bounded above as \begin{align*}
		E\left[\left(\Phat_{t}[j] - p[j]\right)^2\right] \leq \frac{p[j](1-p[j])}{\sum_{i=0}^t n_i\alpha_i^2}.
	\end{align*} Noting that the estimate found at stage $t$ is a valid distribution if $n_t\alpha_{t} - n_t\alpha_{t}^2 \leq \sum_{i=0}^{t-1} n_i\alpha_{i}^2$ (from the condition in Lemma~\ref{lem:upper-bound-one-step}) and $\sum_j p[j](1-p[j]) \leq 1-\sfrac{1}{k}$ leads to result for the $\ell_2^2$ loss. Using the Cauchy-Schwarz inequality leads to the upper bound on the $\ell_1$ loss.
\end{IEEEproof}

\subsection{The unprocessed empirical estimator sequence}
\begin{IEEEproof}[Proof of Corollary \ref{cor:upper-bound-unprocessed}]
	The proof proceeds inductively. For $t=0$, we have $E[\ell_2^2(\Phat_0, p)] = \sfrac{1}{n_0}$. Assume that the estimates at stages $i, 0\leq i\leq t-1$ are the empirical distributions $\phat_{\mathrm{emp}}(X_0^{n_0}, \dots, X_i^{n_i})$ of the samples $X_i^{n_i}$ respectively, and that the error at stage $t-1$ is bounded above by $\left(\sum_j p[j](1-p[j])\right)\left(\sum_{i=0}^{t-1} \frac{n_i}{\left(\sum_{j=0}^i n_j\right)^2}\right)$. It is easy to inductively see that each estimator in this sequence is unbiased by observing that the expected value of the empirical distribution at stage $t$ equals $\alpha_t p + (1-\alpha_t)\Phat_{t-1}$. 
	
	When the estimator at stage $t$ is also the empirical estimator of the samples seen up to batch $t$, we have the upper bound \eqref{eq:unprocessed-final-bound} as follows:
	\begin{align*}
		E[\ell_2^2(\Phat_{t}^{\mathrm{emp}}, p)] =& \sum_{j\in [1:k]}E\left[\Bigg(\frac{\sum_{i=0}^{t-1} n_i}{\sum_{i'=0}^tn_{i'}}\Phat^{\mathrm{emp}}_{t-1}[j] + \frac{n_t}{\sum_{i'=0}^tn_{i'}}\phat_{\mathrm{emp}}(X_t^{n_t})[j] - p[j]\Bigg)^2\right]\\
		=& \sum_{j\in [1:k]} \Biggl(\left(\frac{\sum_{i=0}^{t-1} n_i}{\sum_{i'=0}^tn_{i'}}\right)^2\cdot\var\left(\Phat^{\mathrm{emp}}_{t-1}[j]\right) + \left(\frac{n_t}{\sum_{i'=0}^tn_{i'}}\right)^2\cdot\var\left(\phat_{\mathrm{emp}}(X_t^{n_t})[j]\right)\\
		& \qquad\quad + \frac{n_t\cdot\sum_{i=0}^{t-1} n_i}{\left(\sum_{i'=0}^tn_{i'}\right)^2}\cdot\cov\left(\Phat^{\mathrm{emp}}_{t-1}[j], \phat_{\mathrm{emp}}(X_t^{n_t})[j]\right)\Biggr)\\
		\overset{(a)}{=}& \sum_{j\in [1:k]} \Bigg(\left(\frac{\sum_{i=0}^{t-1} n_i}{\sum_{i'=0}^tn_{i'}}\right)^2\cdot\var\left(\Phat^{\mathrm{emp}}_{t-1}[j]\right) + \frac{n_t\cdot\sum_{i=0}^{t-1} n_i}{\left(\sum_{i'=0}^tn_{i'}\right)^2}\cdot\left(2(1-\alpha)\var\left(\Phat^{\mathrm{emp}}_{t-1}[j]\right)\right) \\
		&\qquad\quad+  \left(\frac{n_t}{\sum_{i'=0}^tn_{i'}}\right)^2\cdot\Bigg(\frac{p[j](1-p[j])}{n_t} +\var\left(\Phat^{\mathrm{emp}}_{t-1}[j]\right)(1-\alpha)^2\left(1-\frac{1}{n_t}\right)\Bigg) \Bigg)\\
		=& \sum_{j\in [1:k]}\Bigg(\var(\Phat_{t-1}^{\mathrm{emp}}[j])\cdot\left(\left(1{-}\frac{\alpha_tn_t}{\sum_{s{=}0}^t n_{s}}\right)^{2} {-} \frac{(1{-}\alpha_t)^2n_t}{\left(\sum_{i{=}0}^tn_i\right)^2}\right) + \frac{p[j](1-p[j])n_t}{\left(\sum_{i=0}^t n_i\right)^2}\Bigg)\\
		\leq& \left(\sum_jp[j](1{-}p[j])\right){\cdot}\left(\sum_{i=0}^{t-1}\frac{n_i}{\left(\sum_{i'=0}^{i}n_{i'}\right)^2} {+} \frac{n_t}{\left(\sum_{i=0}^tn_i\right)^2}\right)\\
		\leq & \left(\sum_jp[j](1{-}p[j])\right)\cdot\left(\sum_{i=0}^{t}\frac{n_i}{\left(\sum_{i'=0}^{i}n_{i'}\right)^2}\right).\numberthis\label{eq:unprocessed-final-bound}
	\end{align*} The expressions for the variance of $\phat^{\mathrm{emp}}(.)$ and its covariance with $\Phat^{\mathrm{emp}}_{t-1}$ used for step $(a)$ are derived in \eqref{eq:empirical-stage-2-variance} and \eqref{eq:covariance-empirical-prev} respectively.
	Taking the supremum of $\sum_jp[j](1-p[j])$ over $p\in \Delta_k$ concludes the proof.
\end{IEEEproof}

\section{Proofs of the oracle-assisted bounds}\label{sec:proofs-oracle}
In this section, we prove the oracle-assisted bounds. Recall that in the oracle-assisted problem, each sample $X_{i}[s]$ is accompanied by source identity information $W_i[s] \sim \mathrm{Bern}(\alpha_{i})$ provided by an oracle, such that the conditional distribution of $X_i[s]$ is
\begin{align*}
	\mathbf{P}\left\{X_i[s] = j\middle|W_i[s]=0, \Pbar_{i-1, (p)}\right\} =& p[j]\\
	\mathbf{P}\left\{X_i[s] = j\middle|W_i[s]=1, \Pbar_{i-1, (p)}\right\} =& \Phat_{i-1}[j].
\end{align*}
\begin{IEEEproof}[Proof of Lemma~\ref{lem:oracle-minimax}]
	Denote the conditional distribution of each sample at stage $i$ given the auxiliary information $W_i^{n_i}$ as $\tilde{P}_{X|W}(p, \Phat_{i-1}, \alpha)$, and for each $i\in [1:n]$, denote the joint distribution of $(X_0^{n_0}, W_1^{n_1}, X_1^{n_1}, \dots, W_i^{n_i}, X_i^{n_i})$ as $\Pbar'_{i, (p)}$. Now for any $p_1, p_2 \in \Delta_k$, for every $i\in [1:t]$,
	\begin{align}
		E\left[D\left(\tilde{P}_{X|W}({p_v, \Phat_{i{-}1}, \alpha_{i}})\middle\|\tilde{P}_{X|W}({p_{v{-}2e_j}, \Phat_{i{-}1}, \alpha_{i}})\right)\right]
		 = \alpha D(p_v\|p_{v{-}2e_j}).\label{eq:oracle-kl-decomposition}
	\end{align}where the expectation is taken with respect to both $P_{W_{i}^{n_{i}}}$ and $ \Pbar'_{i{-}1, (p_v)}$. With the same construction of $\Pc_k$ as Section \ref{sec:proofs-lb}, with the chain rule of the KL divergence along with \eqref{eq:oracle-kl-decomposition}, we find that \begin{align*}
		D\Big(\Pbar'_{t, (p_v)}  \Big\| \Pbar'_{t, (p_{v-2e_j})}\Big) =& D\left(p_v\|p_{v-2e_j}\right)\left(\sum_{i=0}^t n_i\alpha_i\right)
		 \leq \left(\sum_{i=0}^t n_i\alpha_i\right)\frac{8\delta^2}{k(1-\delta^2)}.
	\end{align*} Using Lemma~\ref{lem:lower-bound-partial} with the choice
	\begin{align*}
		\delta^2 = \frac{k}{64\sum_{i=0}^t n_i\alpha_i} <1
	\end{align*} leads to
	\begin{align*}
		r_{t, k}^{\ell_2^2, \mathrm{oracle}} \geq \frac{1}{512\sum_{i=0}^tn_i\alpha_{i}}
	\end{align*}
	whenever $\sum_{i=0}^t n_i\alpha_i \geq \sfrac{k}{4}$. 
	
	For the upper bound, consider the sequence of estimators $\Phat_{t, \mathrm{oracle}}$ that simply outputs the empirical distribution of the real samples accumulated by stage $t$ (i.e. the samples with the corresponding auxiliary $W_i[s] = 0$), and $p_u$ if there are no real samples (i.e. when $W_i[s] = 1$ for every $i, s$). Denote the number of real samples accumulated till stage $t$ as $Z_t$. Conditioned on $Z_t=z>0$, the worst-case expected loss of $\Phat_{t, \mathrm{oracle}}$ is bounded above as
	\begin{align*}
		E\left[\ell_2^2\left(\Phat_{t, \mathrm{oracle}}, p\right)\middle|Z_t=z\right] \leq \frac{1-\frac{1}{k}}{z}.
	\end{align*} Using the trivial upper bound of $\sqrt{2}$ on the $\ell_2^2$ loss between any two elements of $\Delta_k$, we have
	\begin{align}
		E\left[\ell_2^2\left(\Phat_{t, \mathrm{oracle}}, p\right)\right]
		 \leq \sqrt{2}\cdot\Pbar'_{t,({p})}\left\{Z_t < \frac{1}{2}\sum_{i=0}^t n_i\alpha_{i} \right\} + \frac{2\left(1-\frac{1}{k}\right)}{\sum_{i=0}^tn_i\alpha_{i}}. \label{eq:oracle-assisted-ub-incomplete}
	\end{align}Now, $E[Z_t] = \sum_{i=0}^t n_i\alpha_{i}$. Since $Z_t$ is the sum of $\sum_{i=0}^tn_i\alpha_{i}$ independent Bernoulli random variables, using the Hoeffding inequality,
	\begin{align}
		\Pbar'_{t,({p})}\left\{Z_t - \sum_{i=0}^t n_i\alpha_i < -\frac{1}{2}\sum_{i=0}^t n_i\alpha_{i} \right\}
		 \leq \exp\left(-\frac{1}{2}\sum_{i=0}^t n_i\alpha_i\right) \leq \frac{2}{\sum_{i=0}^t n_i\alpha_i}.\label{eq:oracle-assisted-prob-ub}
	\end{align}Combining \eqref{eq:oracle-assisted-ub-incomplete} and \eqref{eq:oracle-assisted-prob-ub} completes the proof.
\end{IEEEproof}

\section{Proof of the Lower Bound for the Non-Matching Regime}\label{app:non-matching}
We start with the following Lemma which will be useful in proving the proposition:
\begin{lemma}\label{lem:recursive-sqrt-sum}
	Let $a_i, i\in [0:\infty)$, be a non-negative sequence such that $a_t \leq M \sum_{i=0}^{t-1} a_i$ for every $t$ and a universal positive constant $M$. Then
	\begin{align}
		\sum_{i=1}^t \frac{a_i}{\sqrt{\sum_{j=0}^{i-1} a_j}} \leq D\sqrt{\sum_{i=0}^t a_i} \label{eq:recursive-sqrt-sum}
	\end{align}for
	\begin{align*}
		D:= 1+ \sqrt{1+\sup_{s\geq 1} \frac{a_s}{\sum_{i=0}^{s-1} a_i}}.
	\end{align*}
\end{lemma}The proof of this Lemma is given in Appendix \ref{app:proofs-lemmas}. We are now ready to prove Proposition~\ref{prop:non-matching}.

\begin{IEEEproof}[Proof of Proposition~\ref{prop:non-matching}]
	The proof proceeds in two parts: \begin{enumerate}
		\item Showing that for $\alpha_t = \Oc\left(\sfrac{1}{\sum_{i=0}^{t-1}n_i\alpha_i^2}\right)$ and $n_t\alpha_{t} = \Oc\left(\sum_{i=0}^{t-1}n_i\alpha_i\right)$,
		\begin{align*}
			\sum_{i=0}^t n_i\alpha_i^2 = \Oc\left({\sqrt{\sum_{i=0}^t n_i\alpha_i}}\right)
		\end{align*}
		\item Showing that when the loss of the estimates at stages $i\in [0:t-1]$ is $\Oc\left(\sfrac{1}{{\sqrt{\sum_{j=0}^{i} n_j\alpha_j}}}\right)$, the error term \begin{align*}
			h_t(\sfrac{1}{4k})=\Oc\left({\sqrt{\sum_{i=0}^t n_i\alpha_i}}\right).
		\end{align*}
	\end{enumerate} The lower bound then follows directly from Theorem~\ref{thm:lower-bound}.
	
	We prove the first part by contradiction. Assume that $\sqrt{\sum_{i=0}^tn_i\alpha_i} = o\left(\sum_{i=0}^t n_i\alpha_i^2\right)$. Then for $\epsilon>0$, there exists $t_\epsilon \geq 0$ such that \begin{align*}
		\frac{1}{\sum_{i=0}^t n_i\alpha_i^2} \leq & \frac{\epsilon}{\sqrt{\sum_{i=0}^t n_i\alpha_i}}&  \forall t\geq t_\epsilon.
	\end{align*}
	Also, since $\alpha_t = \Oc(\sfrac{1}{\sum_{i=0}^{t-1}n_i\alpha_i^2}),$ let $\alpha_t \leq \sfrac{C}{\sum_{i=0}^{t-1}n_i\alpha_i^2}$. Then for all $t\geq t_\epsilon$,\begin{align*}
		\sum_{i=0}^t n_i\alpha_i^2 \leq & C\sum_{i=0}^t \frac{n_i\alpha_i}{\sum_{j=0}^{i-1} n_j\alpha_j^2}
		\leq C_\epsilon + \epsilon C\sum_{i=0}^t \frac{n_i\alpha_i}{\sqrt{\sum_{j=0}^{i-1} n_i\alpha_i}}
		\overset{(a)}{\leq} C_\epsilon + \epsilon CD \sqrt{\sum_{i=0}^t n_i\alpha_i},
	\end{align*}where
	\begin{align*}
		C_\epsilon =& \sum_{i=0}^{t_\epsilon} \frac{n_i\alpha_i}{\sum_{j=0}^{i-1} n_j\alpha_j^2}
		\text{ and } D =\left(1 + \sqrt{1+ \sup_{s\geq 1} \frac{n_s\alpha_s}{\sum_{i=0}^{s-1}n_i\alpha_i}}\right).
	\end{align*}The inequality $(a)$ is a consequence of Lemma~\ref{lem:recursive-sqrt-sum}. We can thus lower bound the limit
	\begin{align*}
		\lim_{t\uparrow\infty} \frac{\sqrt{\sum_{i=0}^t n_i\alpha_i}}{\sum_{i=0}^t n_i\alpha_i^2} \geq& \frac{1}{\epsilon CD + \lim_{t\uparrow\infty} \frac{C_\epsilon}{\sqrt{\sum_{i=0}^t n_i\alpha_i}}}
		 = \frac{1}{\epsilon C D} >0,
	\end{align*}resulting in a contradiction. We thus have 
	\begin{align*}
		\sum_{i=0}^t n_i\alpha_i^2 = \Oc\left({\sqrt{\sum_{i=0}^t n_i\alpha_i}}\right)
	\end{align*}		
	
	We bound the error term as follows:
	\begin{align*}
		\sum_{i=1}^tn_i\alpha_ig_i(\sfrac{1}{4k}) \overset{(b)}{\leq}&  \sum_{i=1}^t n_i\alpha_i\cdot 16k^2 \sup_pE[\ell_2^2(\Phat_{i-1}, p)]\\
		\leq&  16Mk^2 \sum_{i=1}^t \frac{n_i\alpha_i}{\sqrt{\sum_{j=0}^{i-1} n_j\alpha_j}} \\
		\overset{(c)}{\leq}& 16Mk^2D\sqrt{\sum_{i=0}^t n_i\alpha_i},
	\end{align*}where $(b)$ is a consequence of Chebyshev's inequality and $(c)$ follows from Lemma~\ref{lem:recursive-sqrt-sum} for
	\begin{align*}
		D := \left(1 + \sqrt{1+ \sup_{s\geq 1} \frac{n_s\alpha_s}{\sum_{i=0}^{s-1}n_i\alpha_i}}\right).
	\end{align*}
	
\end{IEEEproof}

\section{Proofs of Auxiliary Lemmas}\label{app:proofs-lemmas}
In this section, we present complete proofs for the Lemmas and results used in Sections \ref{sec:proofs-lb}, \ref{sec:proofs-ub}, and \ref{sec:proofs-oracle}.
\begin{IEEEproof}[Proof of Lemma~\ref{lem:lower-bound-partial}]
	We first prove the lemma for the $\ell_2^2$ loss. Define the sign$(\cdot)$ function as\begin{align*}
		\mathrm{sign}(x) = \begin{cases}
			-1 & x<0\\
			0 & x=0\\
			1 & x>0.
		\end{cases}
	\end{align*}and let $\openone\{.\}$ be the indicator function. 
	\begin{align*}
		\sup_{p \in \Delta_k} E_{\Pbar_{t, (p)}}[l_2^2(\Phat_t, p)]
		  \geq& \sup_{p \in \Pc_k}  E_{\Pbar_{t, (p)}}[l_2^2(\Phat_t, p)]\\
		\geq& \sup_{v\in\Vc}  E_{\Pbar_{t, (p_{v})}}\left[\sum_{j=1}^{k/2}\left(\Phat_t - \frac{1}{k} - v_j\frac{\delta}{k}\right)^2\right] \numberthis \label{eq:l-2-fork}\\
		\geq& \frac{\delta^2}{k^2}\sup_{v\in\Vc}E_{\Pbar_{t, (p_{v})}}\left[\sum_{j=1}^{k/2}\openone\left\{\mathrm{sign}\left(\Phat_t - \frac{1}{k}\right) \neq v_j\right\}\right]\numberthis\label{eq:l-2-penalty}\\
		\geq & \frac{\delta^2}{k^2}\sum_{v\in\Vc}\frac{1}{|\Vc|}\sum_{j=1}^{k/2} E_{\Pbar_{t, (p_{v})}}\left[\openone\left\{\mathrm{sign}\left(\Phat_t - \frac{1}{k}\right) \neq v_j\right\}\right]\\
		=& \frac{\delta^2}{k^2} \sum_{j=1}^{k/2}\frac{1}{|\Vc|} \sum_{v\in\Vc} E_{\Pbar_{t, (p_{v})}}\left[\openone\left\{\mathrm{sign}\left(\Phat_t - \frac{1}{k}\right) \neq v_j\right\}\right] \numberthis \label{eq:minimax-self-consuming-pre-mix}
	\end{align*}
	Let $\Vc_{+j} = \{v\in \Vc: v_j = +1\}$ and $\Vc_{-j} = \{v\in \Vc: v_j = -1\}$. Note that these are sets of equal size, and their union is $\Vc$. Denote the mixtures $\frac{1}{|\Vc_{+j}|}\sum_{v\in\Vc_{+j}} \Pbar_{t, (p_{v})}$ and $\frac{1}{|\Vc_{-j}|}\sum_{v\in\Vc_{-j}} \Pbar_{t, (p_{v})}$ as $\Qbar_{t, (+j)}$ and $\Qbar_{t, (-j)}$ respectively. 
	
	Rewriting \eqref{eq:minimax-self-consuming-pre-mix} using $E_Q[\openone\{A\}] = Q(A)$,
	\begin{align*}
		r^{l_2^2}_{\lambda, n_0, n_1}
		 \geq& \frac{\delta^2}{k^2} \sum_{j=1}^{k/2} \frac{1}{2}\Bigg[ \Qbar_{t, (+j)}\left(\mathrm{sign}\left(\Phat_t - \frac{1}{k}\right) = -1 \right)
		+ \Qbar_{t, (-j)}\left(\mathrm{sign}\left(\Phat_t - \frac{1}{k}\right) = +1 \right)\Bigg]\\
		=& \frac{\delta^2}{2k^2}\sum_{j=1}^{k/2} \Bigg[1 - \Qbar_{t, (+j)}\left(\mathrm{sign}\left(\Phat_t - \frac{1}{k}\right) = +1 \right)+ \Qbar_{t, (-j)}\left(\mathrm{sign}\left(\Phat_t - \frac{1}{k}\right) = +1 \right)\Bigg]\\
		\geq & \frac{\delta^2}{2k^2} \sum_{j=1}^{k/2} \left[1- \sup_{A \subset \Xc^{n_0+n_1}}\left( \Qbar_{t, (+j)}(A) - \Qbar_{t, (-j)}(A)\right) \right]\\
		=& \frac{\delta^2}{2k^2} \sum_{j=1}^{k/2}\left[1- \|\Qbar_{t, (+j)} - \Qbar_{t, (-j)}\|_{\mathrm{TV}}\right]\\
		\overset{(a)}{\geq}& \frac{\delta^2}{2k^2} \sum_{j=1}^{k/2}\left[1-\frac{2}{|\Vc|} \sum_{\substack{v\in\Vc:\\v_j=1}} \|\Pbar_{t, (p_{v})} - \Pbar_{t, (p_{v-2e_j})}\|_{\mathrm{TV}}\right]\\
		\geq & \frac{\delta^2}{2k^2} \sum_{j=1}^{k/2}\left[1- \max_{\substack{v\in\Vc:\\v_j=1}} \|\Pbar_{t, (p_{v})} - \Pbar_{t, (p_{v-2e_j})}\|_{\mathrm{TV}}\right]\\
		\geq& \frac{\delta^2}{4k}\left(1 - \frac{2}{k}\sum_{j=1}^{k/2} \max_{\substack{v\in\Vc:\\v_j=1}} \|\Pbar_{t, (p_{v})} - \Pbar_{t, (p_{v-2e_j})}\|_{\mathrm{TV}}\right)\\
		\overset{(b)}{\geq}& \frac{\delta^2}{4k}\left(1 - \sqrt{\frac{2}{k}\sum_{j=1}^{k/2} \max_{\substack{v\in\Vc:\\v_j=1}} \|\Pbar_{t, (p_{v})} - \Pbar_{t, (p_{v-2e_j})}\|_{\mathrm{TV}}^2}\right)\\
		\overset{(c)}{\geq}& \frac{\delta^2}{4k}\left(1 - \sqrt{\frac{2}{k}\sum_{j=1}^{k/2} \max_{\substack{v\in\Vc:\\v_j=1}} D_{KL}\left(\Pbar_{t, (p_{v})} \| \Pbar_{t, (p_{v-2e_j})}\right)}\right) \numberthis\label{eq:lower-bound-tv}
	\end{align*}
	where $(a)$ is due to Jensen's inequality, $(b)$ is due to the Cauchy-Schwarz inequality, and $(c)$ is due to Pinsker's inequality. The result for the $\ell_1$ loss follows directly by observing that for the $\ell_1$ loss, the inequality \eqref{eq:l-2-fork} holds with absolute values in the expectation instead of squares, and the equivalent penalty in \eqref{eq:l-2-penalty} is $\sfrac{\delta}{k}$ instead of $\sfrac{\delta^2}{k^2}$; the rest of the analysis follows exactly the same steps.    
\end{IEEEproof}

\begin{IEEEproof}[Proof of Lemma~\ref{lem:upper-bound-uncombined}]
	First, note that the empirical estimator is the normalized sum of indicator random variables of the form \begin{align*}
		\phat_{\mathrm{emp}}(X_1^{n_1})[j] = \frac{\sum_{s=1}^{n_1}\openone\{X_{1}[s] = j\}}{n_1}.
	\end{align*} Conditioned on $\Phat_{0}[j]$, each of these indicator random variables is distributed as $\mathrm{Bern}\left(\alpha p[j] + (1-\alpha)\Phat_{0}[j]\right)$ independently of the others. If $\Phat_0$ is an unbiased estimate, then $E[\phat_{\mathrm{emp}}(X_1^{n_1})] = E[E[\phat_{\mathrm{emp}}(X_1^{n_1})|\Phat_{0}]] = E[\alpha\Phat_0+(1-\alpha)p] = p$, and therefore, $\phat_{\mathrm{emp}}(X_1^{n_1})$ is an unbiased estimate of $p$. 
	
	Consequently, since the sum of the coefficients of $\phat_{\mathrm{emp}}$ and $\Phat_{1}$ in \eqref{eq:conditionally-unbiased-estimator} is $1$, $\Phat_{1, (\mathrm{cond})}$ is also an unbiased estimate of $p$. Also,
	\begin{align*}
		E&\left[\Phat_{0}[j]\Phat_{1, (\mathrm{cond})}[j]\right]\\
		=& E\left[E\left[\Phat_{0}[j]\Phat_{1, (\mathrm{cond})}[j]\middle|\Phat_{0}[j]\right]\right]\\
		=& E\Bigg[\frac{\Phat_{0}[j]}{\alpha}\Big(\alpha p[j] + (1-\alpha)\Phat_0[j] - (1-\alpha)\Phat_{0}[j]\Big)\Bigg]\\
		=& E\left[p[j]\Phat_{0}[j]\right] = p[j]^2.
	\end{align*}
	This concludes the proof of the first two parts of the Lemma. We now compute the variance of the estimator $\Phat_{1, (\mathrm{cond})}$. The variance of $\Phat_{\mathrm{emp}}[j] := \phat_{\mathrm{emp}}(X_1^{n_1})[j]$ is
	\begin{align*}
		E\left[\left(\Phat_{\mathrm{emp}}[j] - p[j]\right)^2\right]
		 = &  E\left[(\Phat_{\mathrm{emp}}[j])^2\right] - p[j]^2\\
		=&  E\left[E\left[\left(\Phat_{\mathrm{emp}}[j]\right)^2\middle|\Phat_0[j]\right]\right] - p[j]^2\\
		=&  E\Bigg[\left(\alpha p[j] + (1-\alpha)\Phat_0[j]\right)^2\left(1-\frac{1}{n_1}\right) + \frac{1}{n_1}\left(\alpha p[j] + (1-\alpha)\Phat_0[j]\right)\Bigg] - p[j]^2\\
		=&  (1-\alpha)^2\left(1-\frac{1}{n_1}\right)E\left[\left(\Phat_0[j] - p[j]\right)^2\right] + \frac{p[j] - p[j]^2}{n_1}\\
		=& (1-\alpha)^2\left(1-\frac{1}{n_1}\right)\eta \cdot p[j]\left(1-p[j]\right) + \frac{p[j](1-p[j])}{n_1}.\\
		=& p[j](1-p[j])\left(\frac{1}{n_1} + \eta(1-\alpha)^2\left(1-\frac{1}{n_1}\right)\right). \label{eq:empirical-stage-2-variance} \numberthis
	\end{align*}The covariance of $\Phat_{\mathrm{emp}}$ and $\Phat_{0}$ is
	\begin{align*}
		E\Big[\Phat_{\mathrm{emp}}[j]\Phat_{0}[j]\Big] - p[j]^2
		 =& E\left[E\left[\Phat_{\mathrm{emp}}[j]\Phat_{0}[j]\big| \Phat_{0}[j]\right]\right] - p[j]^2\\
		=&  E\left[\left(\alpha p[j] + (1-\alpha)\Phat_{0}[j]\right)\Phat_{0}[j]\right] - p[j]^2\\
		=& \alpha p[j]^2 + (1-\alpha) E\left[\left(\Phat_{0}[j]\right)^2\right] - p[j]^2\\
		=& (1-\alpha) \left(E\left[\left(\Phat_{0}[j]\right)^2\right] - p[j]^2\right)\\
		=& (1-\alpha)\eta \cdot p[j]\left(1-p[j]\right). \numberthis\label{eq:covariance-empirical-prev}
	\end{align*}
	Using \eqref{eq:empirical-stage-2-variance} and \eqref{eq:covariance-empirical-prev}, the variance of $\Phat_{1, (\mathrm{cond})}[j]$ is then
	\begin{align*}
		E\Big[\left(\Phat_{1, (\mathrm{cond})}[j] - p[j]\right)^2\Big] 
		= & \frac{1}{\alpha^2}\Big(\var\left(\Phat_{\mathrm{emp}}[j]\right) + (1-\alpha)^2\var\left(\Phat_{0}[j]\right)
		- 2(1-\alpha)\cov\left(\Phat_{0}[j], \Phat_{\mathrm{emp}}[j]\right)\Big)\\
		=& \frac{p[j]\left(1-p[j]\right)}{\alpha^2}\Big(\frac{1}{n_1} + \eta(1-\alpha)^2\left(1-\frac{1}{n_1}\right) +  (1-\alpha)^2\cdot \eta - 2\eta(1-\alpha)^2\Big)\\
		=& \frac{p[j]\left(1-p[j]\right)}{n_1\alpha^2}\left(1-\eta(1-\alpha)^2\right).
	\end{align*}
\end{IEEEproof}

\begin{IEEEproof}[Proof of Lemma~\ref{lem:recursive-sqrt-sum}] We proceed by induction.
	It is easy to verify \eqref{eq:recursive-sqrt-sum} for $t=1$. Assume that it holds for $t= u-1$. Then
	\begin{align*}
		\sum_{i=1}^u \frac{a_i}{\sqrt{\sum_{j=0}^{i-1} a_j}} \leq& D\sqrt{\sum_{i=0}^{u-1} a_i} + \frac{a_u}{\sqrt{\sum_{i=0}^{u-1} a_i}}\\
		=& \frac{D\sum_{i=0}^{u-1} a_i + a_u}{\sqrt{\sum_{i=0}^{u-1} a_i}}.
	\end{align*}Now
	\begin{align*}
		\frac{D\sum_{i=0}^{u-1} a_i + a_u}{\sqrt{\sum_{i=0}^{u-1} a_i}} \leq& D\sqrt{\sum_{i=0}^u a_i}\\
		\impliedby a_u \leq & D\left(\sqrt{\sum_{i=0}^{u-1} a_i}\sqrt{a_u + \sum_{i=0}^{u-1} a_i} - \sum_{i=0}^{u-1} a_i\right)\\
		\impliedby a_u \leq & D\left(\sum_{i=0}^{u-1} a_i\right)\left(\sqrt{1+ \frac{a_u}{\sum_{i=0}^{u-1} a_i}} - 1\right)\\
		\impliedby \frac{a_u}{\sum_{i=0}^{u-1} a_i} \leq & D\left(\sqrt{1+ \frac{a_u}{\sum_{i=0}^{u-1} a_i}} - 1\right)\\
		\impliedby D\geq & \sqrt{1+ \frac{a_u}{\sum_{i=0}^{u-1} a_i}} + 1.
	\end{align*}Noting that $D$ is the supremum over $u$ of the RHS completes the proof.
\end{IEEEproof}

\bibliographystyle{ieeetr}
\bibliography{ref_arxiv_final}

\vfill

\end{document}